\documentclass[11pt,preprintnumbers,amsmath,amssymb,superscriptaddress]
{revtex4}
\pdfoutput=1
\usepackage[colorlinks=true,urlcolor=blue,linkcolor=blue]{hyperref}
\usepackage{graphicx}
\usepackage{epsfig}
\usepackage{bm}

 \pdfoutput=1 
\usepackage[utf8]{inputenc}
\usepackage{graphicx}
\usepackage{epsfig}
\usepackage{bm}
\usepackage{amssymb}
\usepackage{float}
\usepackage{amsmath}
\usepackage{dcolumn}
\usepackage{cancel}
\usepackage{graphicx}
\usepackage{epsfig}
\usepackage{bm}
\usepackage[greek,english]{babel} 
\usepackage{alphabeta}
\usepackage{amssymb}
\usepackage{float}
\usepackage{amsmath}
\usepackage{dcolumn}
\usepackage{cancel}
 
\def\b{\beta}

 \def\m{\mu}
\def\n{\nu}

\newcommand{\nn}{\nonumber}
 \newcommand{\be}{\begin{equation}}
\newcommand{\ee}{\end{equation}}

\newcommand{\bea}{\begin{eqnarray}}
 
\newcommand{\eea}{\end{eqnarray}}

\def\f{\phi}

\def\z{\zeta}

\usepackage{hyperref}        
\hypersetup{colorlinks=true,urlcolor=blue}

\usepackage[misc]{ifsym}

\begin{document}


\title{Alleviating $H_0$ tension in scalar-tensor and bi-scalar-tensor 
theories }

\author{Maria Petronikolou}
\email{petronikoloumaria@mail.ntua.gr}
 \affiliation{National Observatory of Athens, Lofos Nymfon, 11852 Athens, 
Greece}
\affiliation{Department of Physics, National Technical University of Athens, 
Zografou
Campus GR 157 73, Athens, Greece}

\author{Emmanuel N. Saridakis}
\email{msaridak@noa.gr}
 \affiliation{National Observatory of Athens, Lofos Nymfon, 11852 Athens, 
Greece}
\affiliation{CAS Key Laboratory for Researches in Galaxies and Cosmology, 
Department of Astronomy, University of Science and Technology of China, Hefei, 
Anhui 230026, P.R. China}
 \affiliation{Departamento de Matem\'{a}ticas, Universidad Cat\'{o}lica del 
Norte, 
Avda.
Angamos 0610, Casilla 1280 Antofagasta, Chile}

\begin{abstract}
  We investigate scalar-tensor and bi-scalar-tensor  modified theories of 
gravity that can alleviate the $H_0$ tension.   In the first class of 
theories we show that choosing  particular models with 
shift-symmetric   friction term we are able to alleviate the tension by 
obtaining smaller effective Newton's constant at intermediate times, a feature 
that cannot be easily obtained in modified gravity. In the second 
class of theories,  which involve two extra propagating degrees of freedom, 
we show that the $H_0$ tension can be alleviated, and the 
mechanism behind it is the phantom behavior of the effective dark-energy 
equation-of-state parameter. Hence,  scalar-tensor and bi-scalar-tensor  
  theories  have the capability of alleviating $H_0$ tension with both known
sufficient late-time mechanisms.  
 \end{abstract}

\maketitle

\section{Introduction}
 
Although the Concordance $\Lambda$CDM paradigm of Cosmology, which is 
based on general relativity alongside the cold dark matter sector and the 
cosmological constant, 
 is very successful  in describing the universe evolution, however it seems to 
exhibit possible disadvantages, both at the theoretical, as well as at the 
phenomenological level \cite{Perivolaropoulos:2021jda}. In the first category 
one can find the  cosmological constant problem, as well as the 
non-renormalizability of general 
relativity. In the second category one may find possible cosmological 
tensions.

In particular, a first tension is related to the present value of the Hubble 
parameter $H_0$, since the value estimated by the  
Planck collaboration   is $H_0 = (67.27\pm 0.60)$ km/s/Mpc 
\cite{Planck:2018vyg}, while the direct measurement of the $2019$ SH0ES 
collaboration (R19) gives $H_0 = (74.03\pm1.42)$ km/s/Mpc, namely a tension 
of  about $4.4\sigma$. Furthermore, one has  the $\sigma_8$ 
  related to the matter clustering, and the  possible deviation   of the 
Cosmic Microwave 
Background (CMB) estimation 
\cite{Planck:2018vyg} from the SDSS/BOSS measurement
\cite{Zarrouk:2018vwy,BOSS:2016wmc}. Although there is a big discussion 
whether these tensions are due to   unknown systematics, it seems that   
at least   the  $H_0$ tension  may be indeed   a sign of new physics 
\cite{DiValentino:2021izs, DiValentino:2015ola,  
Bernal:2016gxb,Kumar:2016zpg, 
DiValentino:2017iww,  DiValentino:2017oaw,  Binder:2017lkj, DiValentino:2017zyq,
Sola:2017znb, DEramo:2018vss,Poulin:2018cxd, 
Pan:2019jqh, 
 Pandey:2019plg,Adhikari:2019fvb,
 Perez:2020cwa,Pan:2020bur, 
Benevento:2020fev, Elizalde:2020mfs,Alvarez:2020xmk, Haridasu:2020pms,
Seto:2021xua,Bernal:2021yli,Alestas:2021xes,Krishnan:2021dyb,
Theodoropoulos:2021hkk,
Hu:2015rva,Khosravi:2017hfi,Belgacem:2017cqo,Adil:2021zxp,Nunes:2018xbm,
DiValentino:2019jae,Vagnozzi:2019ezj,
Braglia:2020auw, 
DAgostino:2020dhv, Barker:2020gcp, Wang:2020zfv,Ballardini:2020iws,
LinaresCedeno:2020uxx,daSilva:2020bdc,Odintsov:2020qzd}
(for a review see \cite{Abdalla:2022yfr}).

On the other hand,  modified gravity is a very broad class of theories that aim 
to alleviate the non-renormalizability of general relativity, bypass the 
cosmological constant  problem, and lead to improved cosmological evolution, 
both at the background as well as at the perturbation levels 
\cite{CANTATA:2021ktz,Capozziello:2011et}. In order to construct 
gravitational modifications one can start from the Einstein-Hilbert action of 
 General Relativity and add extra terms in the Lagrangian, resulting to 
  $f(R)$   gravity
\cite{DeFelice:2010aj,Nojiri:2010wj,Starobinsky:2007hu,Cognola:2007zu,
Amendola:2007nt,delaCruz-Dombriz:2006kob,Zhang:2005vt,Faraoni:2007yn,
Basilakos:2013nfa, Papanikolaou:2021uhe},   Gauss-Bonnet and $f(G)$ gravity 
  \cite{Nojiri:2005jg,DeFelice:2008wz,Zhao:2012vta,Shamir:2020ckh}, cubic 
gravity \cite{Asimakis:2022mbe}, Lovelock  gravity
\cite{Lovelock:1971yv,Deruelle:1989fj} etc. Alternatively, one can start form 
the equivalent torsional formulation of gravity and modify it suitably, 
resulting to 
  $f(T)$ gravity 
\cite{Cai:2015emx,Bengochea:2008gz,Linder:2010py,Chen:2010va,Tamanini:2012hg,
Bengochea:2010sg,
Liu:2012fk,Daouda:2012nj,MohseniSadjadi:2012brg,
Finch:2018gkh,Golovnev:2020las,Bejarano:2014bca,
Darabi:2019qpz,Sahlu:2019jmy,Benetti:2020hxp, 
Golovnev:2021htv,Duchaniya:2022rqu}, 
to     
$f(T,T_G)$ gravity 
\cite{Kofinas:2014owa,Kofinas:2014daa}, to $f(T,B)$ gravity
\cite{Bahamonde:2015zma,Farrugia:2018gyz,Escamilla-Rivera:2019ulu,
Caruana:2020szx, Moreira:2021xfe}, etc.
Additionally, one broad class of gravitational modifications is the 
scalar-torsion theories, 
which are constructed by one scalar field coupled to curvature terms. In 
particular,  the    most general  four-dimensional scalar-tensor theory with 
one propagating scalar degree of freedom is Horndeski gravity 
\cite{Horndeski:1974wa} or equivalently generalized Galileon theory 
\cite{DeFelice:2010nf,Deffayet:2011gz,Mota:2010bs,
Barreira:2013eea, Qiu:2011cy,Appleby:2012ba,
Barreira:2014jha,Arroja:2015wpa,Hinterbichler:2015pqa,Babichev:2015rva, 
Brax:2011sv,
Renk:2017rzu}. Finally, one may extend this 
framework to beyond Horndeski 
theories 
\cite{Gleyzes:2014dya,Langlois:2015cwa,Langlois:2018dxi,Babichev:2017guv,
Ilyas:2020qja}, as well as to  bi-scalar-tensor 
theories, in which one has two extra scalar fields  \cite{Naruko:2015zze, 
Saridakis:2016ahq}.

The effect of modified gravity on the late-time universe evolution is two-fold. 
The first is that it induces new terms in the Friedmann equations, which can 
collectively be absorbed into an effective dark-energy sector. The second is 
that it typically leads to a modified Newton's constant.
Hence, in every cosmology governed by a modified theory of gravity one 
typically obtains  Friedmann equations of the form  \cite{CANTATA:2021ktz}
\begin{align}
\label{f11}
H^2 &= \frac{8\pi G_{eff}}{3} \left( \rho_m +\rho_{DE}^{eff} 
\right) ~, 
\\
\dot{H} &= -4\pi G_{eff} \left( \rho_m +\rho_{DE}^{eff} +p_m 
+p_{DE}^{\text{eff}} \right) ~, 
\end{align}
where $\rho_{DE}^{eff}$ and $p_{DE}^{eff}$  are respectively the effective 
dark-energy density and pressure, and $G_{eff}$ is the
effective Newton's constant, all depending on the parameters of the theory.
Hence, qualitatively, 
we deduce that in order to alleviate the $H_0$ tension in this framework, i.e. 
obtain a higher $H_0$ than standard lore predicts, we have two ways 
\cite{Heisenberg:2022lob,Heisenberg:2022gqk}. i) One 
could either try to obtain smaller effective Newton's constant, since 
``weaker'' gravity is reasonable to induce faster expansion. ii) One could  
try to obtain suitable modified-gravity-oriented extra terms in the effective 
dark-energy sector, which could lead to faster expansion, e.g. obtain an 
effective dark-energy equation-of-state parameter $w_{DE}:=p_{DE}/\rho_{DE}$ 
lying in the phantom 
regime. 

In this work we will present two broad classes of modified gravity that 
can fulfill the above qualitative requirements in the correct quantitative way 
and alleviate the $H_0$ tension. The first is scalar-tensor theories
\cite{Horndeski:1974wa}  and the second is  bi-scalar-tensor 
theories     \cite{Naruko:2015zze, Saridakis:2016ahq}. Interestingly enough, we 
show that in the first class the mechanism behind the alleviation is the 
smaller $G_{eff}$, while in the second class it is the phantom dark energy. The 
plan of the work is the following: In Section \ref{scaltens} we briefly review
scalar-tensor theories and we present specific models that can alleviate the 
tension. In Section \ref{biscaltens} we present bi-scalar-tensor theories and 
we construct models  alleviating the tension. Finally, in Section 
\ref{Conclusions} we summarize the obtained results.

\section{Scalar-tensor theories alleviating $H_0$ tension} 
\label{scaltens}

In this section we briefly review scalar-tensor theories and then we present 
particular models that can alleviate the $H_0$ tension.
The most general Lagrangian with one extra scalar 
degree of freedom $\phi$ and curvature terms, giving rise to    
second-order field equations, is $
{\cal L}=\sum_{i=2}^{5}{\cal L}_{i}$
\cite{Horndeski:1974wa,DeFelice:2011bh,Kobayashi:2011nu},
where
\begin{align}
&\!\!\!\!\!\!\!\!\!\!
{\cal L}_{2} = K(\phi,X),\label{eachlag2}\\
&\!\!\!\!\!\!\!\!\!\!{\cal L}_{3} = -G_{3}(\phi,X)\Box\phi,\\
&\!\!\!\!\!\!\!\!\!\!{\cal L}_{4} = G_{4}(\phi,X)\,
R+G_{4,X}\,[(\Box\phi)^{2}-(\nabla_{\mu}\nabla_{\nu}\phi)\,(\nabla^{\mu}
\nabla^{\nu}\phi)]\,,\\
&\!\!\!\!\!\!\!\!\!\!{\cal L}_{5} = G_{5}(\phi,X)\,
G_{\mu\nu}\,(\nabla^{\mu}\nabla^{\nu}\phi)  -\frac{1}{6}\,
G_{5,X}\,[(\Box\phi)^{3}-3(\Box\phi)\,(\nabla_{\mu}\nabla_{\nu}\phi)\,
(\nabla^{\mu}\nabla^{\nu}\phi)
\,\nonumber\\&\ \ \
\ \ \ \ \ \ \ \ \ \ \ \   \ \  \ \ \ \  \ \ \ \ 
 \ \ \ \ \   \ \ \ \  \ \ \ \ \  \ \ \ \ \ \ \ 
+2(\nabla^{\mu}\nabla_{\alpha}\phi)\,(\nabla^
{\alpha}\nabla_{\beta}\phi)\,(\nabla^{\beta}\nabla_{\mu}\phi)]\,.\label{
eachlag5}
\end{align}
As usual,  $R$ is the Ricci scalar, $G_{\mu\nu}$ is  the 
Einstein tensor,   the functions $K$ and $G_{i}$ ($i=3,4,5$) depend on   
$\phi$
and its kinetic energy
$X=-\partial^{\mu}\phi\partial_{\mu}\phi/2$, and
$G_{i,X}:=\partial G_{i}/\partial X$, 
$G_{i,\phi}:=\partial
G_{i}/\partial\phi$.
Focusing on Friedmann-Robertson-Walker (FRW) geometry with metric 
\begin{eqnarray}
 ds^2=-dt^2+a^2(t)\,\delta_{ij} dx^i dx^j,
\end{eqnarray}
 and adding the
 matter Lagrangian ${\cal L}_m$ corresponding to a perfect fluid with energy 
density $\rho_m$  and pressure $p_m$, and performing variation we obtain the 
two 
generalized Friedmann equations: 
\begin{eqnarray}
&&
2X
K_{,X}
-K+6 X\dot{\phi} H G_{3,X}-2X G_{3,\phi}-6H^2 G_{4}
 +24H^2X (G_{4,X}+X G_{4,XX}) -6H \dot{\phi} G_{4,\phi}\nonumber\\
&& -12H X \dot{\phi} G_{4,\phi 
X} +2H^3X \dot{\phi} (5G_{5,X}+2X G_{5,XX}
)   -6H^2X (3G_{5,\phi}+2X G_{5,\phi X})=-\rho_{m}  ,
\label{Fr1gen}
\end{eqnarray}
\begin{eqnarray}
&&K-2X (G_{3,\phi}+\ddot{\phi} G_{3,X})+2(3H^2+2\dot{H}
)G_4
   -8\dot{H}X G_{4,X}
-12H^2 X G_{4,X}-4H\dot{X} G_{4,X}\nonumber\\
&&  
-8H X\dot{X} G_{4,XX}+2(\ddot{\phi}+2H\dot{\phi}) G_{4,\phi}+4X G_{4,
\phi\phi}  +4X (\ddot{\phi}-2H \dot{\phi}) G_{4,\phi X}
\nonumber\\
&&  -4H^2X^2\ddot{\phi} 
G_{5,XX}
-2X (2H^3\dot{\phi}+2H\dot{H} 
\dot{\phi}+3H^2\ddot{\phi}) G_{5,X}  
+4H X(\dot{X}-HX)G_{5,\phi X}\nonumber\\
&&  +4H X\dot{\phi} G_{5,\phi\phi}
+2 [ 2(\dot{H}X+H\dot{X})+3H^2X ] G_{5,\phi}
=-p_m ,
\label{Fr2gen}
\end{eqnarray}
with dots denoting derivatives with respect to $t$. Moreover, variation
 with 
respect to $\phi(t)$ gives
 \begin{equation}
 \label{scalfieldeq}
\frac{1}{a^3}\frac{d}{dt}(a^3J)=P_{\phi}  ,
\end{equation}
with
\begin{eqnarray}
&&
\!\!\!\!\!\!\!\!\!\!\!\!\!\!
J:=
\dot{\phi} K_{,X}+6H X G_{3,X}-2 \dot{\phi} 
G_{3,\phi}-12H X G_{4,\phi 
X} 
+6H^2\dot{\phi}(G_{4,X}+2X G_{4,XX})\nonumber\\
&&  
+2H^3X (3G_{5,X}+2X G_{5,XX}) 
+6H^2\dot{\phi} (G_{5,\phi}+X G_{5,\phi X}),
\end{eqnarray}
\begin{eqnarray}&&P_{\phi}:=
K_{,\phi}-2X (G_{3,\phi\phi}+\ddot{\phi} G_{3,\phi 
X})+6 (2H^2+\dot{H}) G_{4,\phi}\nonumber\\
&&\ \ \ \ \ \ \    
+6H (\dot{X}+2H X) G_{4,\phi X}   -6H^2X G_{5,\phi\phi}+2H^3X \dot{\phi} 
G_{5,\phi 
X} .
\end{eqnarray} 
Lastly, as usual we consider   the 
matter conservation equation
$
\dot{\rho}_m+3H(\rho_m+p_m)  =  0$. 

In the following   we   present specific models of scalar-tensor theories that 
can alleviate $H_0$ tension  \cite{Petronikolou:2021shp}.  Since   
  Horndeski theory recovers $\Lambda$CDM cosmology for
$G_4=1/(16\pi G)$, $K=-2\Lambda=const$, and $G_3=G_5=0$,  our strategy is   
 to introduce  deviations which are negligible at high redshifts, where the
CMB structure is formed, but    become significant at low redshifts, in which 
local Hubble measurements take place.

 We start by  examining a   subclass of Horndeski gravity that 
contains the $G_5$ term, which is called ``non-minimal derivative coupling''. 
In particular, we can consider models with  $G_4=1/(16\pi G)$ and 
$G_3=0$, which is the case in $\Lambda$CDM cosmology, and     impose a 
simple scalar-field potential and standard kinetic term, namely  $K= 
-V(\phi) + X$. Additionally,  since   
 $G_5$   affects the   friction 
terms of the scalar field 
\cite{Saridakis:2010mf, Koutsoumbas:2017fxp,  Karydas:2021wmx}, 
 we  choose  the $G_5$ term to depend only on $X$, i.e. $G_5(\phi,X)=G_5(X)$.
   Inserting into (\ref{Fr1gen}),(\ref{Fr2gen}) gives 
 the effective dark energy  density
and pressure   \cite{Petronikolou:2021shp}
\begin{eqnarray}
\label{rhode}
&&
\!\!\!\!\!\!\!\!\!\!
\rho_{DE}= 2X-K+2H^3X\dot{\phi}(5G_{5,X}+2X G_{5,XX}) ,\\
\label{pde}
&&\!\!\!\!\!\!\!\!\!\! p_{DE}= K-2X G_{5,X}\left(
2H^3\dot{\phi}+2H\dot{H}\dot{\phi}+3H^2 \ddot{\phi}\right) 
 -4H^2 
X^2\ddot{\phi}G_{5,XX},
 \end{eqnarray} 
 and thus the dark-energy equation-of-state parameter becomes
 \begin{eqnarray}
\label{wde1}
w_{DE}\equiv\frac{p_{DE}}{\rho_{DE}}. 
 \end{eqnarray}

One can  choose $G_5(X)$ suitably   in order for   $H(z)$ 
   to coincide with  $
 H_{\Lambda \text{CDM}}(z) :=H_0 
\sqrt{\Omega_{m0}(1+z)^3+1-\Omega_{m0}}$   at $z= z_{\rm 
CMB}\approx 1100$,
namely $H(z\rightarrow z_{\rm CMB}) \approx H_{\Lambda\text{CDM}}(z\rightarrow 
z_{\rm CMB})$,
but satisfy 
$H(z\rightarrow 0) > H_{\Lambda\text{CDM}}(z\rightarrow 0)$.  For simplicity, 
 we   focus on   dust matter (i.e. $p_m=0$), and without loss of generality we 
consider $K= -V_0\phi + X$.

We start with the investigation of a model with   
\begin{eqnarray}
 G_5(X)=\xi X^{2}.
\end{eqnarray}
In this 
case (\ref{rhode}) and (\ref{pde})  give
\begin{eqnarray}
\label{rhode2}
&&
\!\!\!\!\!\!\!\!\!\!\!\!\!\!\!\!
\rho_{DE}=  \frac{\dot{\phi}^2}{2}+V_0\phi+7\xi H^3\dot{\phi}^5,\\
\label{pde2}
&&\!\!\!\!\!\!\!\!\!\!
\!\!\!\!\!\!p_{DE}=  \frac{\dot{\phi}^2}{2}-V_0\phi-\xi\dot{\phi}^4 \left( 
2H^3\dot{\phi}+2H\dot{H}\dot{\phi}+5H^2\ddot{\phi}\right).
 \end{eqnarray}
 
\begin{figure} [H]
\centering
\includegraphics[width=0.65\textwidth]{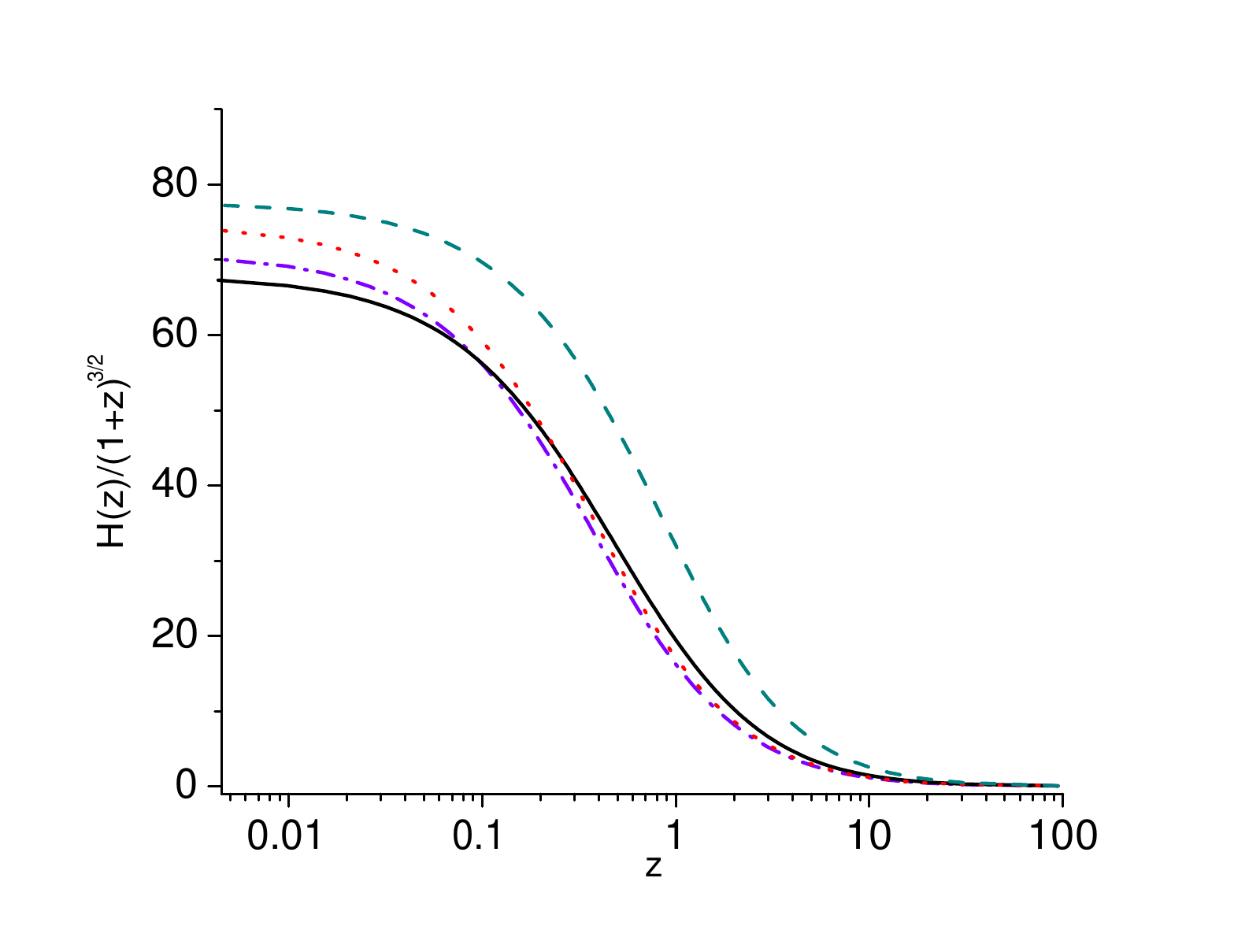} \vspace{-0.5cm}
	\caption{{\it{  The normalized $H(z)/(1+z)^{3/2}$  in 
units of km/s/Mpc  as a function of the redshift, for $\Lambda$CDM cosmology 
(black - solid) and for  scalar-tensor  Model I  with $V_0=0.06$   and with 
$G_5(X)=\xi 
X^{2}$, for $\xi=1.7$ 
(purple - dashed-dotted),  $\xi=1.4$ 
(red - dotted)
 and  $\xi=1.1$ (blue - dashed), in  $H_0$  units. We have imposed 
$\Omega_{m0}\approx0.31$.}}  
}
	\label{xiX2}
\end{figure}

As we mentioned above we chose the model parameter $V_0$ and the initial 
conditions for the scalar field in order to obtain $H(z_{\rm CMB}) 
=H_{\Lambda\text{CDM}}( z_{\rm CMB})$ and $\Omega_{m0}=0.31$ in agreement with 
 \cite{Planck:2018vyg}, and we handle $\xi$ as the parameter that determines 
the late-time deviation from $\Lambda$CDM cosmology. 
 We solve the Friedmann equation numerically and in  Fig.  \ref{xiX2} we depict 
$H(z)/(1+z)^{3/2}$ for  
different choices of $\xi$. 
As we see,   the model coincides with $\Lambda$CDM one at high 
and intermediate redshifts, while at small redshifts it leads to
higher values of $H_0$. In particular,  $H_0$ depends on 
the model 
parameter $\xi$, and it can be around $H_0 \approx 74$ 
km/s/Mpc for $\xi=1.3$ (we mention here  that since $\xi$ 
has dimensions of $[M]^{-9}$  and since
$H_0\approx 10^{-61}$ in Planck  units, this gives    $\xi^{1/9}\sim 
10^{40}$GeV$^{-1}$).
Hence, we can see that  $H_0$ tension can be alleviated at 3$\sigma$ if 
$1.2<\xi<1.7$. 

\begin{figure}[H]
\vspace{-5.5cm}
	\includegraphics[width=0.5\textwidth]{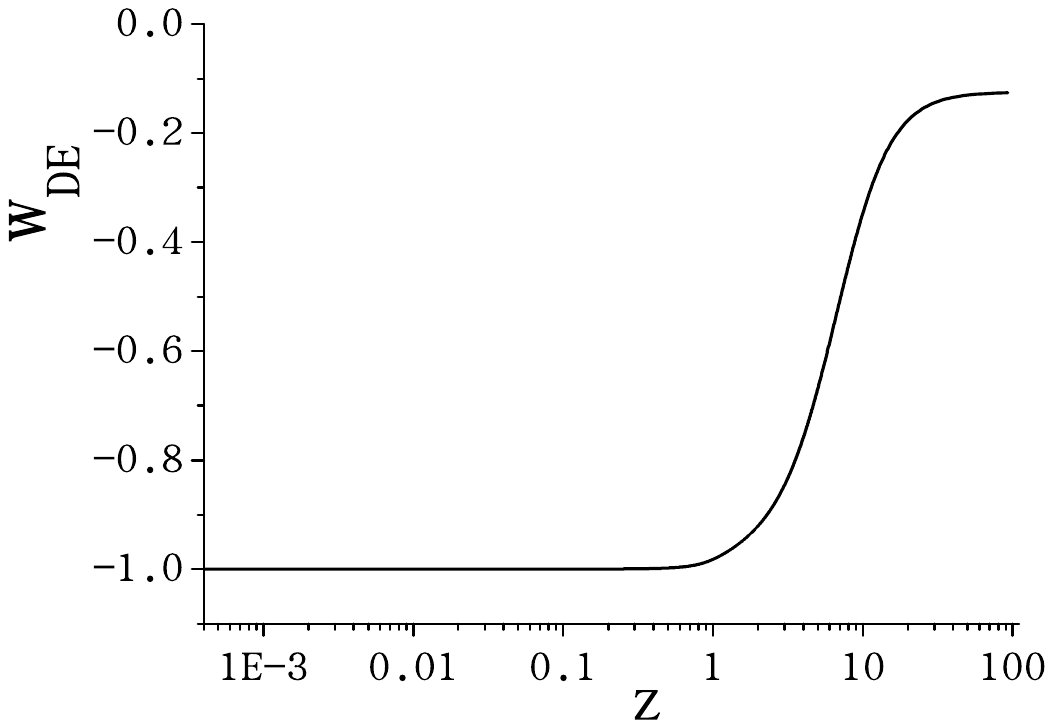}
	\includegraphics[ width=0.5\textwidth]{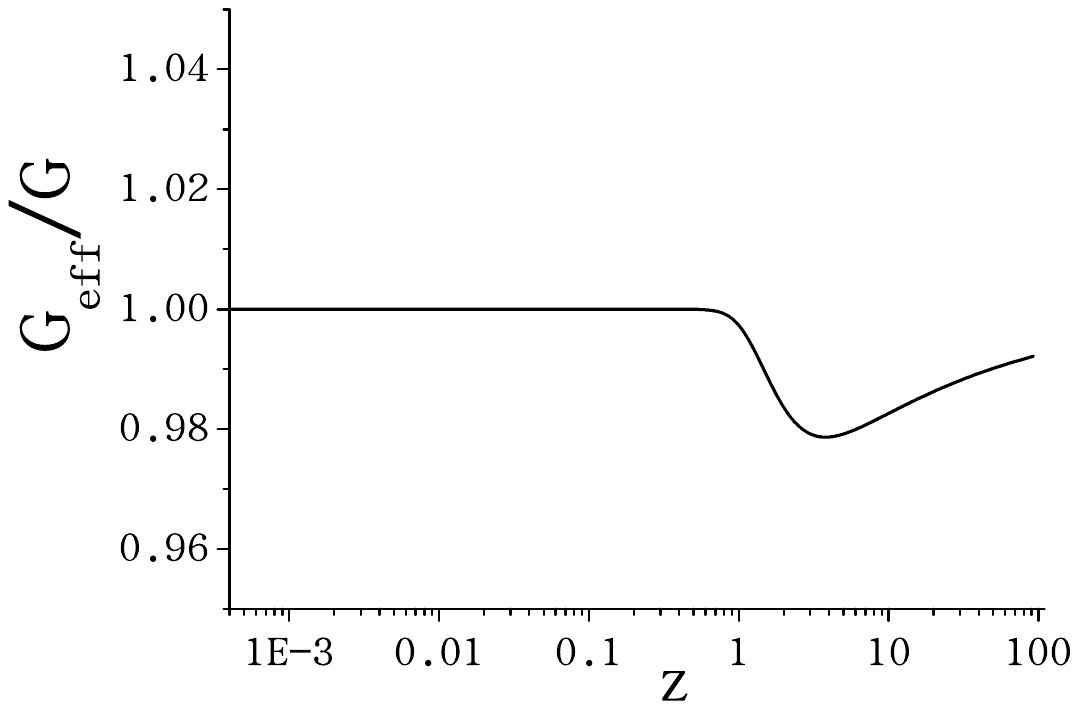}
	\caption{{\it{ {\bf{ Left Graph}}:
The  effective dark-energy equation-of-state parameter $ 
w_{DE}$ given in 
  (\ref{wde1})  as a function of the 
redshift, for Model I   with $V_0=0.08$  and with $\xi=1.3$ in  $H_0$  
units. 	 	 {\bf{Right graph}}: The corresponding normalized 
effective Newton's constant $ 
\frac{G_{eff}}{G}$ given in 
(\ref{GeffG}) as a function of the 
redshift. The graphs are from \cite{Petronikolou:2021shp}}}.
}
	\label{wDEModelI}
\end{figure}

Let us now examine what is the mechanism behind the tension alleviation, 
following \cite{Petronikolou:2021shp}.
In the left graph of    Fig. \ref{wDEModelI} we depict the  effective dark-energy 
equation-of-state parameter $ 
w_{DE}$ given in 
  (\ref{wde1}). As we can see it does not exhibit phantom behavior, namely it 
cannot be the cause of increased $H_0$ 
\cite{Heisenberg:2022lob,Heisenberg:2022gqk}). On the other hand, we remind 
that in scalar-tensor Horndeski gravity, one obtains an effective Newton's 
constant given by 
  \cite{Bellini:2014fua,Peirone:2017ywi}
\begin{equation}
\frac{G_{eff}}{G}=\frac{1}{2}\left(G_4-2X G_{4,X}+X 
G_{5,\phi}-\dot{\phi}HXG_{5,X}
 \right)^{-1}.
\label{GeffG}
\end{equation}
 In the right graph of Fig. \ref{wDEModelI} we depict the evolution of the 
normalized 
effective Newton's constant $\frac{G_{eff}}{G}$. As  we can see, we obtain a 
decrease of the effective
Newton's constant at intermediate redshifts, and as we mentioned in the 
Introduction this can lead to an increased $H_0$. Hence, we deduce that in the 
scenario at hand, the mechanism behind the tension alleviation is the decreased 
$G_{eff}$, namely suitably weaker gravity.

Finally, let us discuss the perturbative behavior of the model at hand. As one 
can show 
\cite{DeFelice:2010pv,DeFelice:2011bh,Appleby:2011aa}, in order  for 
Horndeski/generalized Galileon theory to be 
free from Laplacian instabilities
associated with the scalar field propagation speed one should have
\begin{equation}
c_{S}^{2}\equiv\frac{3(2w_{1}^{2}w_{2}H-w_{2}^{2}w_{4}+4w_{1}w_{2}\dot{w}_{1}
-2w_{1}^{2}
\dot{w}_{2})
}{w_{1}(4w_{1}w_{3}+9w_{2}^{2})}
\geq0,
\label{cscon}
\end{equation}
while  in order not to have perturbative ghosts one should have
  \begin{equation}
Q_{S}\equiv\frac{w_{1}(4w_{1}w_{3}+9w_{2}^{2})}{3w_{2}^{2}}>0 .
\label{Qscon}
\end{equation}
Additionally,  the light speed in these theories is 
 \cite{DeFelice:2011bh}
\begin{equation}
c_{T}^{2}\equiv\frac{w_4 }{w_1}
\geq0,
\label{cTcon2}
\end{equation}
which at late times should be very close to 1, in agreement with LIGO/Virgo 
   bounds \cite{Ezquiaga:2017ekz}. By studying $c_{S}^{2}$,  $Q_{S}$ and 
$c_{T}^{2}$ one can show that the scenario at hand is viable 
\cite{Petronikolou:2021shp}, although with an amount of fine-tuning.

We   mention that we could examine other models which can 
lead to similar behavior, namely a smaller effective Newton's constant due to 
the friction term $G_5(X)$ that can result to higher $H_0$.  For instance a 
model with  $G_5(X)=\lambda X^{4}$
  also leads to  
  $H_0 \approx 74$ 
km/s/Mpc for $\lambda=1$ in  $H_0$  units (since $\lambda$ has dimensions of 
$[M]^{-17}$ we acquire
$\lambda^{1/17}\sim 10^{30}$GeV$^{-1}$),
and   the tension can 
be alleviated at 3$\sigma$ if 
$0.5<\lambda<1.2$, in  $H_0$  units. On the other hand, one can see that models 
with odd powers of $X$ do not solve the tension, since the last term in 
(\ref{GeffG}) changes signs and this does not guarantee that $G_{eff}/G$ 
will 
remain smaller than 1.

\begin{figure}[H]
    \centering
	\includegraphics[width=0.6\textwidth]{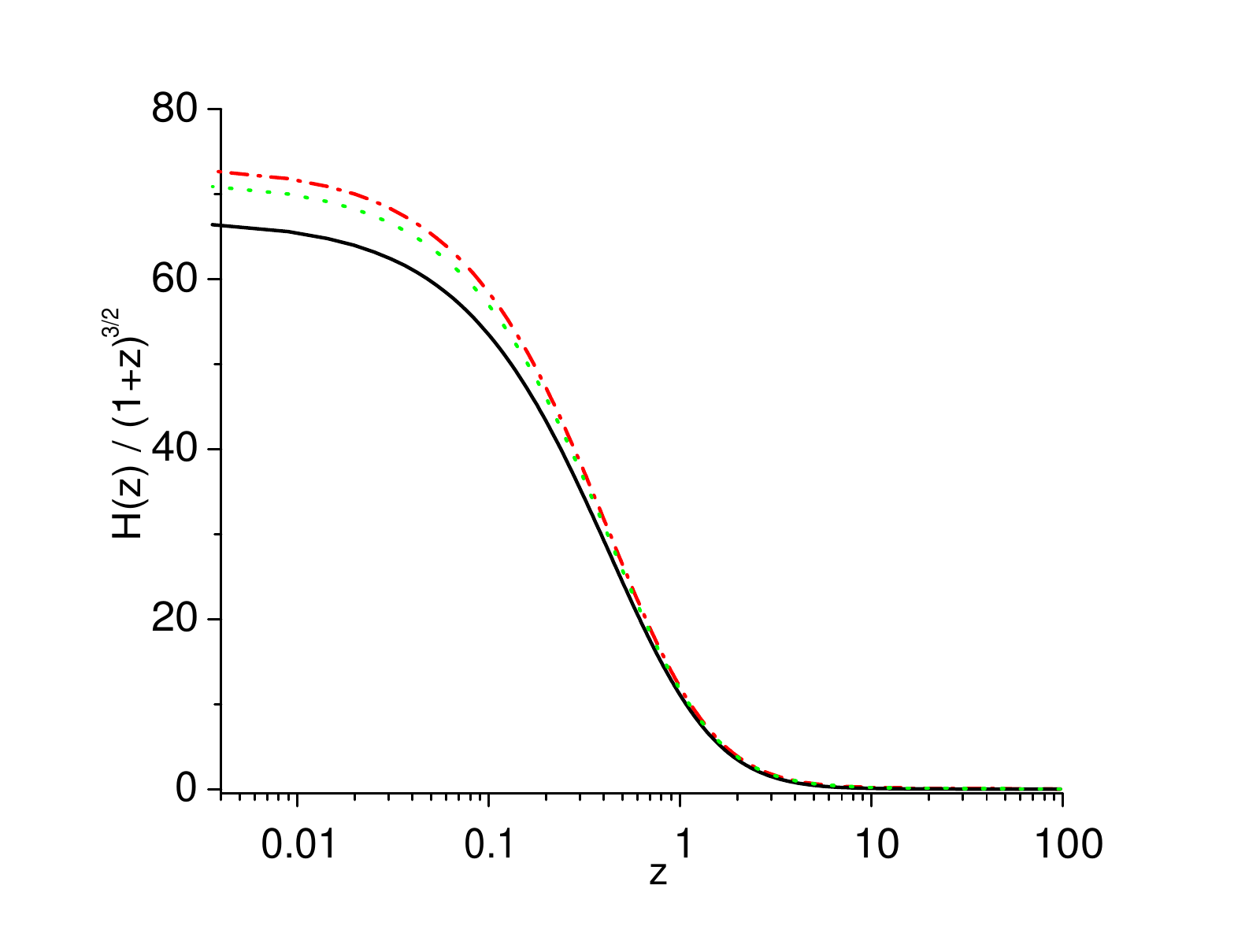}
	\caption{{\it{  The normalized $H(z)/(1+z)^{3/2}$  in 
units of km/s/Mpc  as a function of the redshift, for $\Lambda$CDM cosmology 
(black - solid) and for  the combined scalar-tensor model (\ref{combmod}) with 
$V_0=0.08$   
and with $\xi=1.5$, $\lambda=0.001$ 
(green - dotted), 
 and  $\xi=1.3$, $\lambda=0.002$  (red - dashed-dotted), in  $H_0$  units. 
We have imposed  $\Omega_{m0}\approx0.31$. 
}}}
	\label{wDEModelIII}
\end{figure}

Finally, we can consider combinations of monomial  forms, such as  
\begin{equation}
 G_5(X)=\xi X^{2}+\lambda X^{4},\label{combmod}
\end{equation}
in which case we have more freedom to obtain 
the desired decreased $G_{eff}/G$. We elaborate the equations numerically, and 
in Fig. \ref{wDEModelIII}  we present the normalized Hubble constant evolution.
As we observe, the $H_0$ tension is alleviated due to the decreased effective 
Newtons' constant.  

We close this section mentioning that   although in modified theories of 
gravity in general one acquires an effective Newton's constant different than 
the standard one,   $G_{eff}/G<1$ in a viable way is not easily 
obtained. For instance, in    $f(R)$ gravity where
\cite{Basilakos:2013nfa}
\begin{eqnarray} 
\frac {G_{eff}}{G}=\frac{1}{f_{R}}\frac{1+4\frac{k^{2}}{a^{2}} 
\frac{f_{RR}}{f_{R}}}
{1+3\frac{k^{2}}{a^{2}} \frac{f_{RR}}{f_{R}}},
\end{eqnarray}
with $k$ the wave-number, one can see that under the viability conditions 
$f_{,R}>0$  for $R\geq R_0$ (with $R_0$ the present value of the Ricci scalar) 
 and 
$ f_{,RR}>0$ for 
$R\geq R_0$ \cite{DeFelice:2010aj}, as well as  
$ 0<\frac{R f_{,RR}}{f_{,R}}(r)<1 $ at $r=-\frac{Rf_{,R}}{f}=-2$,
 \cite{DeFelice:2010aj}  $G_{eff}/G<1$ cannot be obtained. On the other hand, 
this is indeed possible in $f(T)$ gravity, where 
$G_{eff}=G\,\left(1+\frac{\partial f(T)}{\partial T}\right)^{-1}$ 
\cite{Nesseris:2013jea}.   However, scalar-tensor theories may present such 
behavior quite easily.

In summary, as we see,  the above particular sub-class of 
scalar-tensor gravity can alleviate the $H_0$ tension due to 
the effect of the kinetic-energy-dependent $G_5$-term on decreasing $G_{eff}$.

\section{Bi-scalar-tensor theories alleviating $H_0$ tension}
\label{biscaltens}
 
In the section we present another class of modified gravity that can lead to 
the alleviation of $H_0$ tension, namely bi-scalar theories of gravity.
These theories are determined by the 
 action   \cite{Naruko:2015zze, 
Saridakis:2016ahq}
\begin{equation}
\label{bfR}
S=\int d^{4}\sqrt{-g}\, f\left(R,(\nabla R)^{2},\square R \right),
\end{equation}
with $(\nabla R)^{2}=g^{\m\n}\nabla_{\m}R\nabla_{\n}R$. In this work we focus 
 on models with $ 
f(R,(\nabla R)^{2},\square R)=\mathcal{K}((R,(\nabla 
R)^{2})+\mathcal{G}(R,(\nabla 
R)^{2})\square R $.
We can   rewrite the above action  by transforming   the Lagrangian  using 
double Lagrange multipliers, in which case one can clearly see that they 
correspond to bi-scalar-tensor theories of gravity.
Hence, introducing the scalar fields  $\phi$ and $\chi$  
  through 
$g_{\mu\nu}=\frac{1}{2}e^{-\sqrt{\frac{2}{3}}\chi}\hat{g}_{\mu\nu}$  and
$\varphi:=f_{\b}$, with $\beta:=\square R$, we obtain \cite{Banerjee:2022ynv}
 \begin{eqnarray}
\label{action}
&&\!\!\!\!\!\!\!\!\!\!\!\!\!\!\!
S=\int d^{4}x 
\sqrt{-\hat{g}}\left[\frac{1}{2}\hat{R}-\frac{1}{2}\hat{g}^{\m\n}\nabla_{\m}
\chi\nabla_{\n
}\chi\right. 
- 
\frac{1}{\sqrt{6}}e^{-\sqrt{\frac{2}{3}}\chi}\hat{g}^{\m\n}\mathcal{G}\nabla_{\m
}
\chi\nabla_{\n}\phi+\frac{1}{4}e^{
-2\sqrt{
\frac{2}{3}}
\chi}\mathcal{K}
\nonumber\\
&& \ \ \ \ \ \ \ \ \ \ \ \ \ \ \ \ 
\left.
+\frac{1}{2}e^{-\sqrt{\frac{2}{3}}\chi}\mathcal{G}\hat{\square}\phi-\frac{1}{4}
e^{-\sqrt{
\frac{2}{3}}\chi}\phi\right].
\end{eqnarray}
Thus, varying the above action in terms of the metric, we extract 
 the 
Friedmann equations as \cite{Naruko:2015zze, Saridakis:2016ahq} 
\begin{eqnarray}
&&H^2=\frac{1}{3}(\rho_{DE}+\rho_m)
 \label{FR1b}
 \\
&&2\dot{H}+3H^2=-(p_{DE}+p_m),
 \label{FR2b}
\end{eqnarray}
 where we have defined an effective dark-energy sector with energy density  
and pressure given by 
\begin{eqnarray}
  \label{rhoDE}
 &&\!\!\!\!\!\!\!\!\!\!\!\!\!\!\!\!\!\!\!\!
 \rho_{DE}:=
 \frac{1}{2}\dot{\chi}^{2}-\frac{1}{4}e^{-2\sqrt{\frac{2}{
3}}\chi}\mathcal{K} 
-\frac{2}{3}\dot{\phi}^{2}\left[\dot{\phi}\left(\sqrt{6}\dot{\chi}
-9H\right)-3\ddot{\phi} 
\right]\mathcal{G}_{B}
\nn\\
&& 
+\frac{1}{2}e^{-\sqrt{\frac{2}{3}}\chi}\left[\dot{B}\dot{\f}\mathcal{
G}_{B}+ \frac{\f}{2} +\dot{\f}^{2}\left(\mathcal{G}_{\f}-2\mathcal{K}_{B} 
\right)
 \right],
 \end{eqnarray}
\begin{eqnarray}
\label{pDE}
&&\!\!\!\!\!\!\!\!\!\!\!\!\!\!\!\!\!\!\!\!\!\!\!\!\!\!\!\!\!\!\!\!\!\!\!\!\!\!\!
\!\!
p_{DE}:= 
 \frac{1}{2}\dot{\chi}^{2}+\frac{1}{4}e^{-2\sqrt{
\frac{2}{3}}
\chi}\mathcal{K} 
+
\frac{1}{2} e^{-\sqrt{\frac{2}{3}}\chi}\left(\dot{B}\dot{\f
}\mathcal{
G}_{B}+\dot{\f}^{2}\mathcal{G}_{\f}-\frac{\f}{2} \right),
\end{eqnarray}  
with  
$
\mathcal{K}=\mathcal{K}(\phi,B)$ and $\mathcal{G}=\mathcal{G}(\phi,B),
$
 with
$
B=2e^{\sqrt{
\frac{2}{3}}\chi}g^{\m\n}\nabla_{\m}\phi\nabla_{\n}\phi,
$
and where for simplicity we set the Planck mass to 1.
Moreover, varying the action with respect to the scalar fields, we obtain
their evolution equation as  \cite{Naruko:2015zze, Saridakis:2016ahq} :
\begin{eqnarray}
&&\!\!\!\!\!\!\!\!\!\!\!\!\!\!\!\!\!\!\!\!\!\!\!\!\!
 \ddot{\chi}+3H\dot{\chi}-\frac{1}{3}\dot{\f}^{2}\left[\dot{\f
}
\left(3\sqrt{6}H-
2\dot{\chi} \right)+\sqrt{6}\ddot{\f} \right]\mathcal{G}_{B}
+\frac{1}{\sqrt{6}}e^{-2\sqrt{\frac{2}{3}}\chi}\mathcal{K}\nn\\
&&+
\frac{1}{
2\sqrt{6}}
e^{-\sqrt{\frac{2}{3}}\chi}\left[2\dot{B}\dot{\f}\mathcal{G}_{B}-\f+2\dot{\f}^{2
}
\left(\mathcal{K}_{
B}+\mathcal{G}_{\f} \right) \right] 
=0,
\label{chiequation}
\end{eqnarray}
and
\begin{eqnarray}
&&\!\!\!\!\!\!\!\!\! \!
\frac{1}{3}e^{-\sqrt{\frac{2}{3}}\chi}
\left[\dot{\f}\left(-9H+\sqrt{6}\dot{\chi}
\right)-3\ddot{\f}\right]\mathcal{K}_{B}
 +\frac{1}{6}\dot{B}\left\{3e^{-\sqrt{\frac{2}{3}}
\chi}\dot{B} +4\dot{\f}
\left[\dot{\f}\left(9H-\sqrt{6}\dot{\chi}\right)+3\ddot{\f}\right]\right\}
\mathcal{G}_{BB}
\nn\\
&&\!\!\!\!\!\!\!\!\! \!
+\frac{1}{3}e^{-\sqrt{
\frac{2}{3}
}\chi}\left[\dot{\f}\left(9H-\sqrt{6}\dot{\chi}\right)+3\ddot{\f}\right]\mathcal
{G}_{\f}
+\left\{e^{-\sqrt{\frac{2}{3}}\chi}\dot{B}\dot{\f}+\frac{2}{3}\dot{\f}^{2}
\left[\dot{\f}\left(9H-\sqrt{6}\dot{\chi}\right)+3\ddot{\f}\right]\right\}
\mathcal{G}_{B 
\f}
\nn
\\ 
&&\!\!\!\!\!\!\!\!\! \!
+\left[
 \frac{4}{3}\dot{\f}\left(9H-2\sqrt{6}\dot
{\chi} 
\right)\ddot{\f}
-\frac{1}{\sqrt{6}}e^{
-\sqrt{\frac{2}{3}}\chi}\dot{B}\dot{\chi}
+\dot{\f}^2\left(18H^{2}+6\dot{H}-3\sqrt{6}H\dot{\chi}-\frac{2}{3}\dot{\chi}^{2}
-\sqrt{6} \ddot{\chi}\right)\right]
\mathcal{G}_{B}
\nn
\\
&&\!\!\!\!\!\!\!\!\! \!
-e^{-\sqrt{\frac{2}{3}}\chi}\dot{\f}^{2}\mathcal{K}_{B 
\f}+\frac{1}{2}e^{-\sqrt{\frac{2}{3}}\chi}\dot{\f}^{2}\mathcal{G}_{\f\f}
 -e^{-\sqrt{\frac{2}{3}}\chi}\dot{B}\dot{\f}\mathcal{K}_{BB}
-\frac{1}{4}
e^{-2\sqrt{\frac{2}{3}}\chi}\mathcal{K}_{\f}+ 
\frac{1}{4}e^{-\sqrt{\frac{2}{3}}\chi}
=0,
\label{phiequation}
\end{eqnarray}
with   $\mathcal{G}_{B \f}=\mathcal{G}_{\f 
B}\equiv\frac{\partial^{2}\mathcal{G}}{\partial B \partial \f}$, etc.  
Finally, one can define 
the effective dark-energy equation-of-state parameter as $w_{DE}:=
p_{DE}/\rho_{DE}$. 
  
\begin{figure}[H]
\begin{center}
\includegraphics[height=5.15cm,width=8.6cm, 
clip=true]{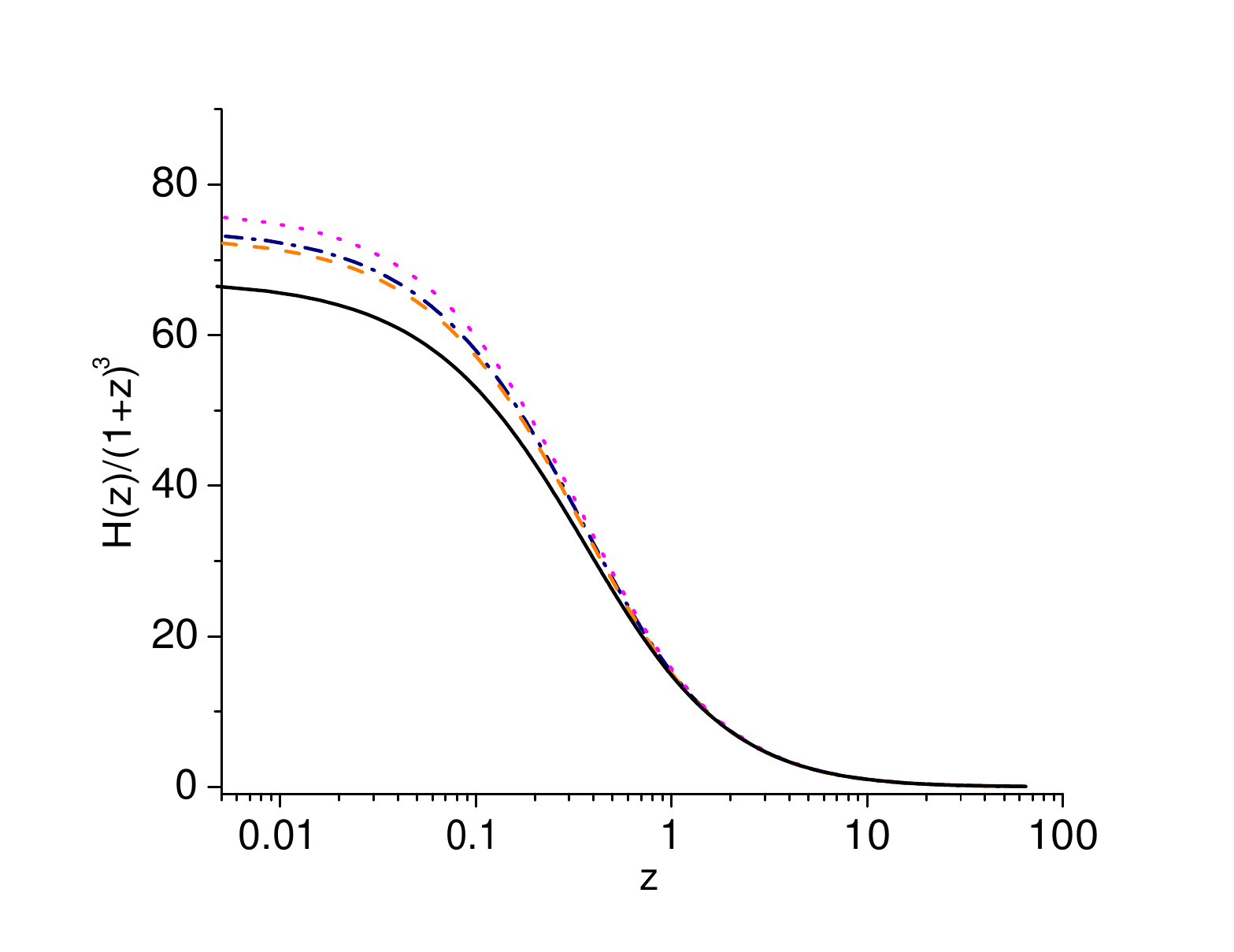}
\includegraphics[height=5.15cm,width=7.7cm, clip=true]{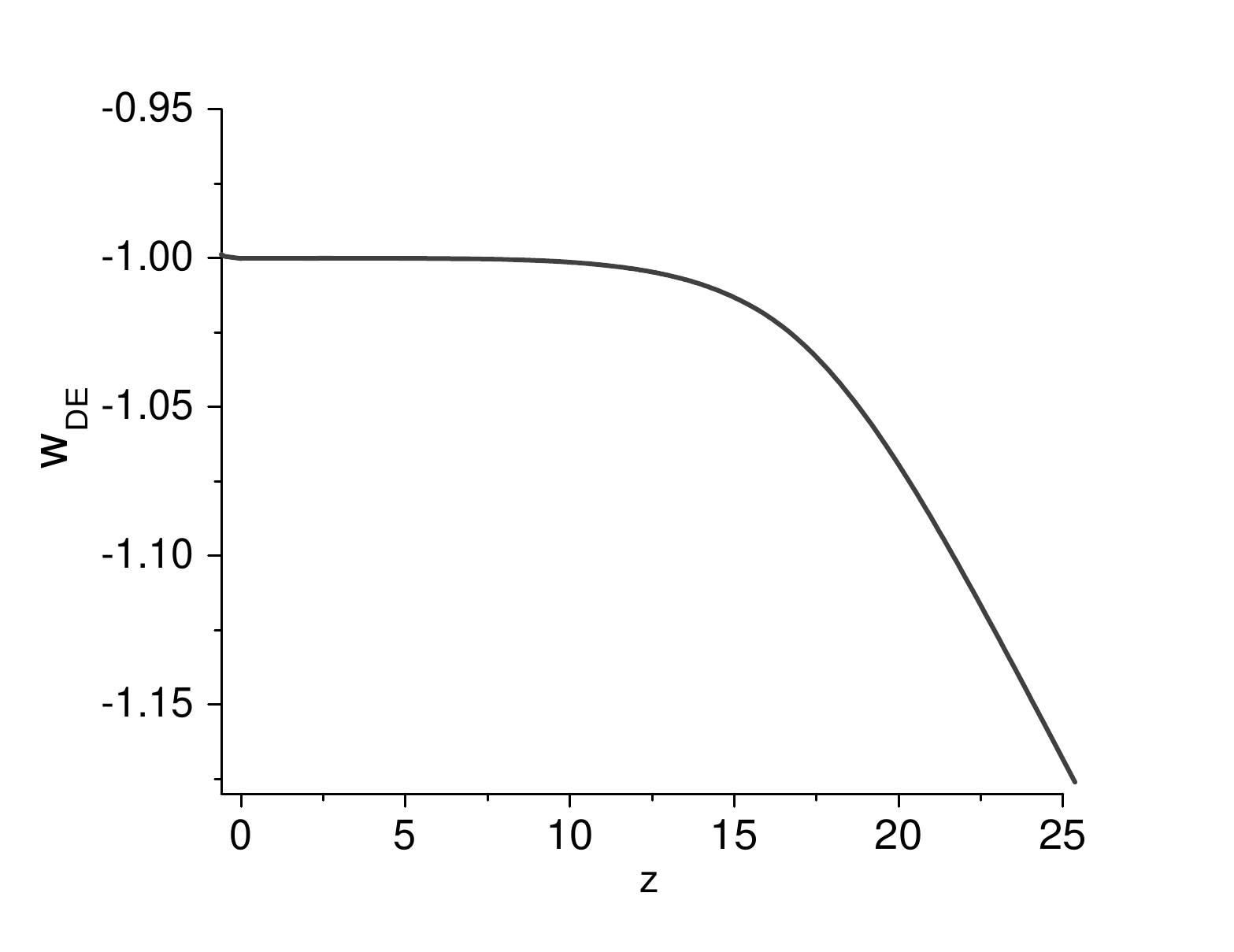}
\caption{{\it {
{\bf{Left Graph}}: The normalized $H(z)/(1+z)^{3}$  in 
units of km/s/Mpc  as a function of the redshift, for $\Lambda$CDM cosmology 
(black solid) and for bi-scalar-tensor  Model I  with
$\zeta=-8$ (dashed 
orange),  with $\zeta=-10$ (dashed-dotted blue), and with
$\zeta=-12$ (dotted magenta), in  Planck units. 
We have imposed 
$\Omega_{m0}\approx0.31$. {\bf{Right graph}}: The corresponding effective 
dark-energy equation-of-state parameter $w_{DE}$ as 
a function of the redshift, for  $\zeta=-10$  in  Planck  units. }}
 } 
\label{H1}
\end{center}
\end{figure}

Let us now extract specific models which   coincide 
with $\Lambda$CDM cosmology at CMB redshifts, while at low-redshifts   deviate 
from it, giving rise to higher $H_0$. 
 A first model that we can examine is Model I, having
 \begin{eqnarray}
  \mathcal{K}(\f,B)=\frac{1}{2}\f-\frac{\z}{2}B \quad \text{and} \quad 
\mathcal{G}(\f,B)=0.
 \end{eqnarray}
 In this case,  
(\ref{rhoDE}),(\ref{pDE}) give
\begin{equation}
  \label{rhoDE2}
 \rho_{DE}= 
 \frac{1}{2}\dot{\chi}^{2}-\frac{1}{8}e^{-2\sqrt{\frac{2}{3}}\chi}\f+\frac{1}{4}
e^{-\sqrt{\frac{2}{3}}\chi}\left(\f+\z\dot{\f}^{2} \right),
 \end{equation}
\begin{equation}
\label{pDE2}
p_{DE}=
 \frac{1}{2}\dot{\chi}^{2}+\frac{1}{8}e^{-2\sqrt{\frac{2}{3}}\chi}\f-\frac{1}{4}
e^{-\sqrt{\frac{2}{3}}\chi}\left(\f-\z\dot{\f}^{2} \right).
\end{equation}
We solve the cosmological equations numerically and in 
 the left graph of Fig. \ref{H1}  we depict the normalised   combination 
$H(z)/(1+z)^{3/2}$ as a function of the redshift  for $\Lambda$CDM cosmology, 
and for Model I for different values of $\zeta$. We find that $H_0$ 
depends on the model parameter $\zeta$ as expected, and for  $\zeta=-10$   it 
is around $H_0\approx 74 km/s/Mpc$, which 
is consistent with its direct measurement    
(note that in natural units  this corresponds to a typical value 
$\zeta^{1/4}\sim-10^{-19}$ GeV$^{-1}$).

Let us now study what is the mechanism behind the $H_0$ alleviation.
In the right graph of Fig. \ref{H1} we present the evolution of  $w_{DE}(z)$, 
and as we observe  most of the time it lies in the phantom regime which, as we 
discussed in the Introduction, is one of the ways one can obtain the tension 
alleviation. Hence, contrary to the case of single-scalar-tensor theories of the 
previous section, where a decreased $G_{eff}$ was the cause of the tension 
alleviation, in the present bi-scalar theories it is the phantom behavior that 
leads to higher $H_0$.

We could proceed to the investigation of other models within the examined 
class. For instance, we can examine  Model II, characterized by
 \begin{eqnarray}
 \mathcal{K}(\f,B)=\frac{1}{2}\f \quad \text{and} \quad 
\mathcal{G}(\f,B)=\xi B.
 \end{eqnarray}
Repeating the same steps as in the previous model    we find 
that  the present Hubble value  $H_0$ depends on the model parameter $\xi$. 
In particular,  for
  $\xi=-10$ it is around 
$H_0\approx 74 km/s/Mpc$ (in natural units   $\xi\sim-10$ 
corresponds to a typical value $\xi^{1/8}\sim-10^{-19}$ GeV$^{-1}$). Similarly 
to the previous case, the mechanism behind  the alleviation is the 
phantom behavior. Hence, we conclude that bi-scalar-tensor theories are very 
efficient in alleviating the $H_0$ tension.

 \section{Conclusions}
 \label{Conclusions}
 
 We investigated scalar-tensor and bi-scalar-tensor  modified theories of 
gravity that can alleviate the $H_0$ tension.   In general,
gravitational modifications affect  the late-time evolution of the universe 
through the new terms they bring in the Friedmann equations, namely in the 
effective dark-energy sector, as well as through the effective Newton's 
constant they induce. If these effects lead to weaker gravity (smaller 
$G_{eff}$) at suitable redshifts, or to more repulsive effective dark-energy 
(for instance exhibiting phantom behavior) then they can cause faster expansion 
comparing to $\Lambda$CDM paradigm, and thus lead to an increased $H_0$ value.

As a first class of models we examined the scalar-tensor theories, namely 
Horndeski/Generalized Galileon  gravity. Choosing  particular models with 
shift-symmetric $G_5$ friction term we were able to alleviate the tension by 
obtaining smaller effective Newton's constant at intermediate times, a feature 
that cannot be easily obtained in modified gravity theories. Additionally, we 
showed that the models at hand are free from perturbative instabilities, and they 
can have gravitational-wave speed equal to the speed of light, nevertheless with 
an amount of fine-tuning.

As a second class we examined bi-scalar-tensor theories, namely 
theories involving two extra propagating degrees of freedom. Choosing 
particular models we showed that the $H_0$ tension can be alleviated, and the 
mechanism behind it is the phantom behavior of the effective dark-energy 
equation-of-state parameter.

In summary, as we see, scalar-tensor theories with one or two scalar fields 
have the capability of alleviating $H_0$ tension with both sufficient 
mechanisms. Such capabilities may be added in the other known 
phenomenological advantages of these theories, and act as 
additional indication that they could be good candidates for the description of 
Nature.

\subsection*{Acknowledgements}
 M.P. is supported by the Basic Research program
of the National Technical University of Athens (NTUA,
PEVE) 65232600-ACT-MTG: Alleviating Cosmological
Tensions Through Modified Theories of Gravity. The authors acknowledge the 
contribution of the LISA CosWG, and of   COST Actions  CA18108 
``Quantum Gravity 
Phenomenology in the multi-messenger approach''  and 
CA21136 ``Addressing observational tensions in cosmology with systematics and 
fundamental physics (CosmoVerse)''.

\bibliographystyle{utphys} 

\bibliography{SaridakisRefs}

\providecommand{\href}[2]{#2}\begingroup\raggedright\begin{thebibliography}{100}

\bibitem{Perivolaropoulos:2021jda}
L.~Perivolaropoulos and F.~Skara, ``{Challenges for \ensuremath{\Lambda}CDM: An
  update},'' \href{http://dx.doi.org/10.1016/j.newar.2022.101659}{{\em New
  Astron. Rev.} {\bfseries 95} (2022) 101659},
  \href{http://arxiv.org/abs/2105.05208}{{\ttfamily arXiv:2105.05208
  [astro-ph.CO]}}.

\bibitem{Planck:2018vyg}
{\bfseries Planck} Collaboration, N.~Aghanim {\em et~al.}, ``{Planck 2018
  results. VI. Cosmological parameters},''
  \href{http://dx.doi.org/10.1051/0004-6361/201833910}{{\em Astron. Astrophys.}
  {\bfseries 641} (2020) A6}, \href{http://arxiv.org/abs/1807.06209}{{\ttfamily
  arXiv:1807.06209 [astro-ph.CO]}}. [Erratum: Astron.Astrophys. 652, C4
  (2021)].

\bibitem{Zarrouk:2018vwy}
P.~Zarrouk {\em et~al.}, ``{The clustering of the SDSS-IV extended Baryon
  Oscillation Spectroscopic Survey DR14 quasar sample: measurement of the
  growth rate of structure from the anisotropic correlation function between
  redshift 0.8 and 2.2},'' \href{http://dx.doi.org/10.1093/mnras/sty506}{{\em
  Mon. Not. Roy. Astron. Soc.} {\bfseries 477} no.~2, (2018) 1639--1663},
  \href{http://arxiv.org/abs/1801.03062}{{\ttfamily arXiv:1801.03062
  [astro-ph.CO]}}.

\bibitem{BOSS:2016wmc}
{\bfseries BOSS} Collaboration, S.~Alam {\em et~al.}, ``{The clustering of
  galaxies in the completed SDSS-III Baryon Oscillation Spectroscopic Survey:
  cosmological analysis of the DR12 galaxy sample},''
  \href{http://dx.doi.org/10.1093/mnras/stx721}{{\em Mon. Not. Roy. Astron.
  Soc.} {\bfseries 470} no.~3, (2017) 2617--2652},
  \href{http://arxiv.org/abs/1607.03155}{{\ttfamily arXiv:1607.03155
  [astro-ph.CO]}}.

\bibitem{DiValentino:2021izs}
E.~Di~Valentino, O.~Mena, S.~Pan, L.~Visinelli, W.~Yang, A.~Melchiorri, D.~F.
  Mota, A.~G. Riess, and J.~Silk, ``{In the realm of the Hubble
  tension\textemdash{}a review of solutions},''
  \href{http://dx.doi.org/10.1088/1361-6382/ac086d}{{\em Class. Quant. Grav.}
  {\bfseries 38} no.~15, (2021) 153001},
  \href{http://arxiv.org/abs/2103.01183}{{\ttfamily arXiv:2103.01183
  [astro-ph.CO]}}.

\bibitem{DiValentino:2015ola}
E.~Di~Valentino, A.~Melchiorri, and J.~Silk, ``{Beyond six parameters:
  extending $\Lambda$CDM},''
  \href{http://dx.doi.org/10.1103/PhysRevD.92.121302}{{\em Phys. Rev. D}
  {\bfseries 92} no.~12, (2015) 121302},
  \href{http://arxiv.org/abs/1507.06646}{{\ttfamily arXiv:1507.06646
  [astro-ph.CO]}}.

\bibitem{Bernal:2016gxb}
J.~L. Bernal, L.~Verde, and A.~G. Riess, ``{The trouble with $H_0$},''
  \href{http://dx.doi.org/10.1088/1475-7516/2016/10/019}{{\em JCAP} {\bfseries
  10} (2016) 019}, \href{http://arxiv.org/abs/1607.05617}{{\ttfamily
  arXiv:1607.05617 [astro-ph.CO]}}.

\bibitem{Kumar:2016zpg}
S.~Kumar and R.~C. Nunes, ``{Probing the interaction between dark matter and
  dark energy in the presence of massive neutrinos},''
  \href{http://dx.doi.org/10.1103/PhysRevD.94.123511}{{\em Phys. Rev. D}
  {\bfseries 94} no.~12, (2016) 123511},
  \href{http://arxiv.org/abs/1608.02454}{{\ttfamily arXiv:1608.02454
  [astro-ph.CO]}}.

\bibitem{DiValentino:2017iww}
E.~Di~Valentino, A.~Melchiorri, and O.~Mena, ``{Can interacting dark energy
  solve the $H_0$ tension?},''
  \href{http://dx.doi.org/10.1103/PhysRevD.96.043503}{{\em Phys. Rev. D}
  {\bfseries 96} no.~4, (2017) 043503},
  \href{http://arxiv.org/abs/1704.08342}{{\ttfamily arXiv:1704.08342
  [astro-ph.CO]}}.

\bibitem{DiValentino:2017oaw}
E.~Di~Valentino, C.~B\o{}ehm, E.~Hivon, and F.~R. Bouchet, ``{Reducing the
  $H_0$ and $\sigma_8$ tensions with Dark Matter-neutrino interactions},''
  \href{http://dx.doi.org/10.1103/PhysRevD.97.043513}{{\em Phys. Rev. D}
  {\bfseries 97} no.~4, (2018) 043513},
  \href{http://arxiv.org/abs/1710.02559}{{\ttfamily arXiv:1710.02559
  [astro-ph.CO]}}.

\bibitem{Binder:2017lkj}
T.~Binder, M.~Gustafsson, A.~Kamada, S.~M.~R. Sandner, and M.~Wiesner,
  ``{Reannihilation of self-interacting dark matter},''
  \href{http://dx.doi.org/10.1103/PhysRevD.97.123004}{{\em Phys. Rev. D}
  {\bfseries 97} no.~12, (2018) 123004},
  \href{http://arxiv.org/abs/1712.01246}{{\ttfamily arXiv:1712.01246
  [astro-ph.CO]}}.

\bibitem{DiValentino:2017zyq}
E.~Di~Valentino, A.~Melchiorri, E.~V. Linder, and J.~Silk, ``{Constraining Dark
  Energy Dynamics in Extended Parameter Space},''
  \href{http://dx.doi.org/10.1103/PhysRevD.96.023523}{{\em Phys. Rev. D}
  {\bfseries 96} no.~2, (2017) 023523},
  \href{http://arxiv.org/abs/1704.00762}{{\ttfamily arXiv:1704.00762
  [astro-ph.CO]}}.

\bibitem{Sola:2017znb}
J.~Sol\`a, A.~G\'omez-Valent, and J.~de~Cruz~P\'erez, ``{The $H_0$ tension in
  light of vacuum dynamics in the Universe},''
  \href{http://dx.doi.org/10.1016/j.physletb.2017.09.073}{{\em Phys. Lett. B}
  {\bfseries 774} (2017) 317--324},
  \href{http://arxiv.org/abs/1705.06723}{{\ttfamily arXiv:1705.06723
  [astro-ph.CO]}}.

\bibitem{DEramo:2018vss}
F.~D'Eramo, R.~Z. Ferreira, A.~Notari, and J.~L. Bernal, ``{Hot Axions and the
  $H_0$ tension},'' \href{http://dx.doi.org/10.1088/1475-7516/2018/11/014}{{\em
  JCAP} {\bfseries 11} (2018) 014},
  \href{http://arxiv.org/abs/1808.07430}{{\ttfamily arXiv:1808.07430
  [hep-ph]}}.

\bibitem{Poulin:2018cxd}
V.~Poulin, T.~L. Smith, T.~Karwal, and M.~Kamionkowski, ``{Early Dark Energy
  Can Resolve The Hubble Tension},''
  \href{http://dx.doi.org/10.1103/PhysRevLett.122.221301}{{\em Phys. Rev.
  Lett.} {\bfseries 122} no.~22, (2019) 221301},
  \href{http://arxiv.org/abs/1811.04083}{{\ttfamily arXiv:1811.04083
  [astro-ph.CO]}}.

\bibitem{Pan:2019jqh}
S.~Pan, W.~Yang, C.~Singha, and E.~N. Saridakis, ``{Observational constraints
  on sign-changeable interaction models and alleviation of the $H_0$
  tension},'' \href{http://dx.doi.org/10.1103/PhysRevD.100.083539}{{\em Phys.
  Rev. D} {\bfseries 100} no.~8, (2019) 083539},
  \href{http://arxiv.org/abs/1903.10969}{{\ttfamily arXiv:1903.10969
  [astro-ph.CO]}}.

\bibitem{Pandey:2019plg}
K.~L. Pandey, T.~Karwal, and S.~Das, ``{Alleviating the $H_0$ and $\sigma_8$
  anomalies with a decaying dark matter model},''
  \href{http://dx.doi.org/10.1088/1475-7516/2020/07/026}{{\em JCAP} {\bfseries
  07} (2020) 026}, \href{http://arxiv.org/abs/1902.10636}{{\ttfamily
  arXiv:1902.10636 [astro-ph.CO]}}.

\bibitem{Adhikari:2019fvb}
S.~Adhikari and D.~Huterer, ``{Super-CMB fluctuations and the Hubble
  tension},'' \href{http://dx.doi.org/10.1016/j.dark.2020.100539}{{\em Phys.
  Dark Univ.} {\bfseries 28} (2020) 100539},
  \href{http://arxiv.org/abs/1905.02278}{{\ttfamily arXiv:1905.02278
  [astro-ph.CO]}}.

\bibitem{Perez:2020cwa}
A.~Perez, D.~Sudarsky, and E.~Wilson-Ewing, ``{Resolving the $H_0$ tension with
  diffusion},'' \href{http://dx.doi.org/10.1007/s10714-020-02781-0}{{\em Gen.
  Rel. Grav.} {\bfseries 53} no.~1, (2021) 7},
  \href{http://arxiv.org/abs/2001.07536}{{\ttfamily arXiv:2001.07536
  [astro-ph.CO]}}.

\bibitem{Pan:2020bur}
S.~Pan, W.~Yang, and A.~Paliathanasis, ``{Non-linear interacting cosmological
  models after Planck 2018 legacy release and the $H_0$ tension},''
  \href{http://dx.doi.org/10.1093/mnras/staa213}{{\em Mon. Not. Roy. Astron.
  Soc.} {\bfseries 493} no.~3, (2020) 3114--3131},
  \href{http://arxiv.org/abs/2002.03408}{{\ttfamily arXiv:2002.03408
  [astro-ph.CO]}}.

\bibitem{Benevento:2020fev}
G.~Benevento, W.~Hu, and M.~Raveri, ``{Can Late Dark Energy Transitions Raise
  the Hubble constant?},''
  \href{http://dx.doi.org/10.1103/PhysRevD.101.103517}{{\em Phys. Rev. D}
  {\bfseries 101} no.~10, (2020) 103517},
  \href{http://arxiv.org/abs/2002.11707}{{\ttfamily arXiv:2002.11707
  [astro-ph.CO]}}.

\bibitem{Elizalde:2020mfs}
E.~Elizalde, M.~Khurshudyan, S.~D. Odintsov, and R.~Myrzakulov, ``{Analysis of
  the $H_0$ tension problem in the Universe with viscous dark fluid},''
  \href{http://dx.doi.org/10.1103/PhysRevD.102.123501}{{\em Phys. Rev. D}
  {\bfseries 102} no.~12, (2020) 123501},
  \href{http://arxiv.org/abs/2006.01879}{{\ttfamily arXiv:2006.01879 [gr-qc]}}.

\bibitem{Alvarez:2020xmk}
P.~D. Alvarez, B.~Koch, C.~Laporte, and A.~Rinc\'on, ``{Can scale-dependent
  cosmology alleviate the $H_0$ tension?},''
  \href{http://dx.doi.org/10.1088/1475-7516/2021/06/019}{{\em JCAP} {\bfseries
  06} (2021) 019}, \href{http://arxiv.org/abs/2009.02311}{{\ttfamily
  arXiv:2009.02311 [gr-qc]}}.

\bibitem{Haridasu:2020pms}
B.~S. Haridasu, M.~Viel, and N.~Vittorio, ``{Sources of $H_0$-tension in dark
  energy scenarios},''
  \href{http://dx.doi.org/10.1103/PhysRevD.103.063539}{{\em Phys. Rev. D}
  {\bfseries 103} no.~6, (2021) 063539},
  \href{http://arxiv.org/abs/2012.10324}{{\ttfamily arXiv:2012.10324
  [astro-ph.CO]}}.

\bibitem{Seto:2021xua}
O.~Seto and Y.~Toda, ``{Comparing early dark energy and extra radiation
  solutions to the Hubble tension with BBN},''
  \href{http://dx.doi.org/10.1103/PhysRevD.103.123501}{{\em Phys. Rev. D}
  {\bfseries 103} no.~12, (2021) 123501},
  \href{http://arxiv.org/abs/2101.03740}{{\ttfamily arXiv:2101.03740
  [astro-ph.CO]}}.

\bibitem{Bernal:2021yli}
J.~L. Bernal, L.~Verde, R.~Jimenez, M.~Kamionkowski, D.~Valcin, and B.~D.
  Wandelt, ``{The trouble beyond $H_0$ and the new cosmic triangles},''
  \href{http://dx.doi.org/10.1103/PhysRevD.103.103533}{{\em Phys. Rev. D}
  {\bfseries 103} no.~10, (2021) 103533},
  \href{http://arxiv.org/abs/2102.05066}{{\ttfamily arXiv:2102.05066
  [astro-ph.CO]}}.

\bibitem{Alestas:2021xes}
G.~Alestas and L.~Perivolaropoulos, ``{Late-time approaches to the Hubble
  tension deforming H(z), worsen the growth tension},''
  \href{http://dx.doi.org/10.1093/mnras/stab1070}{{\em Mon. Not. Roy. Astron.
  Soc.} {\bfseries 504} no.~3, (2021) 3956--3962},
  \href{http://arxiv.org/abs/2103.04045}{{\ttfamily arXiv:2103.04045
  [astro-ph.CO]}}.

\bibitem{Krishnan:2021dyb}
C.~Krishnan, R.~Mohayaee, E.~O. Colg\'ain, M.~M. Sheikh-Jabbari, and L.~Yin,
  ``{Does Hubble tension signal a breakdown in FLRW cosmology?},''
  \href{http://dx.doi.org/10.1088/1361-6382/ac1a81}{{\em Class. Quant. Grav.}
  {\bfseries 38} no.~18, (2021) 184001},
  \href{http://arxiv.org/abs/2105.09790}{{\ttfamily arXiv:2105.09790
  [astro-ph.CO]}}.

\bibitem{Theodoropoulos:2021hkk}
A.~Theodoropoulos and L.~Perivolaropoulos, ``{The Hubble Tension, the M Crisis
  of Late Time H(z) Deformation Models and the Reconstruction of Quintessence
  Lagrangians},'' \href{http://dx.doi.org/10.3390/universe7080300}{{\em
  Universe} {\bfseries 7} no.~8, (2021) 300},
  \href{http://arxiv.org/abs/2109.06256}{{\ttfamily arXiv:2109.06256
  [astro-ph.CO]}}.

\bibitem{Hu:2015rva}
B.~Hu and M.~Raveri, ``{Can modified gravity models reconcile the tension
  between the CMB anisotropy and lensing maps in Planck-like observations?},''
  \href{http://dx.doi.org/10.1103/PhysRevD.91.123515}{{\em Phys. Rev. D}
  {\bfseries 91} no.~12, (2015) 123515},
  \href{http://arxiv.org/abs/1502.06599}{{\ttfamily arXiv:1502.06599
  [astro-ph.CO]}}.

\bibitem{Khosravi:2017hfi}
N.~Khosravi, S.~Baghram, N.~Afshordi, and N.~Altamirano, ``{$H_0$ tension as a
  hint for a transition in gravitational theory},''
  \href{http://dx.doi.org/10.1103/PhysRevD.99.103526}{{\em Phys. Rev. D}
  {\bfseries 99} no.~10, (2019) 103526},
  \href{http://arxiv.org/abs/1710.09366}{{\ttfamily arXiv:1710.09366
  [astro-ph.CO]}}.

\bibitem{Belgacem:2017cqo}
E.~Belgacem, Y.~Dirian, S.~Foffa, and M.~Maggiore, ``{Nonlocal gravity.
  Conceptual aspects and cosmological predictions},''
  \href{http://dx.doi.org/10.1088/1475-7516/2018/03/002}{{\em JCAP} {\bfseries
  03} (2018) 002}, \href{http://arxiv.org/abs/1712.07066}{{\ttfamily
  arXiv:1712.07066 [hep-th]}}.

\bibitem{Adil:2021zxp}
S.~A. Adil, M.~R. Gangopadhyay, M.~Sami, and M.~K. Sharma, ``{Late-time
  acceleration due to a generic modification of gravity and the Hubble
  tension},'' \href{http://dx.doi.org/10.1103/PhysRevD.104.103534}{{\em Phys.
  Rev. D} {\bfseries 104} no.~10, (2021) 103534},
  \href{http://arxiv.org/abs/2106.03093}{{\ttfamily arXiv:2106.03093
  [astro-ph.CO]}}.

\bibitem{Nunes:2018xbm}
R.~C. Nunes, ``{Structure formation in $f(T)$ gravity and a solution for $H_0$
  tension},'' \href{http://dx.doi.org/10.1088/1475-7516/2018/05/052}{{\em JCAP}
  {\bfseries 05} (2018) 052}, \href{http://arxiv.org/abs/1802.02281}{{\ttfamily
  arXiv:1802.02281 [gr-qc]}}.

\bibitem{DiValentino:2019jae}
E.~Di~Valentino, A.~Melchiorri, O.~Mena, and S.~Vagnozzi, ``{Nonminimal dark
  sector physics and cosmological tensions},''
  \href{http://dx.doi.org/10.1103/PhysRevD.101.063502}{{\em Phys. Rev. D}
  {\bfseries 101} no.~6, (2020) 063502},
  \href{http://arxiv.org/abs/1910.09853}{{\ttfamily arXiv:1910.09853
  [astro-ph.CO]}}.

\bibitem{Vagnozzi:2019ezj}
S.~Vagnozzi, ``{New physics in light of the $H_0$ tension: An alternative
  view},'' \href{http://dx.doi.org/10.1103/PhysRevD.102.023518}{{\em Phys. Rev.
  D} {\bfseries 102} no.~2, (2020) 023518},
  \href{http://arxiv.org/abs/1907.07569}{{\ttfamily arXiv:1907.07569
  [astro-ph.CO]}}.

\bibitem{Braglia:2020auw}
M.~Braglia, M.~Ballardini, F.~Finelli, and K.~Koyama, ``{Early modified gravity
  in light of the $H_0$ tension and LSS data},''
  \href{http://dx.doi.org/10.1103/PhysRevD.103.043528}{{\em Phys. Rev. D}
  {\bfseries 103} no.~4, (2021) 043528},
  \href{http://arxiv.org/abs/2011.12934}{{\ttfamily arXiv:2011.12934
  [astro-ph.CO]}}.

\bibitem{DAgostino:2020dhv}
R.~D'Agostino and R.~C. Nunes, ``{Measurements of $H_0$ in modified gravity
  theories: The role of lensed quasars in the late-time Universe},''
  \href{http://dx.doi.org/10.1103/PhysRevD.101.103505}{{\em Phys. Rev. D}
  {\bfseries 101} no.~10, (2020) 103505},
  \href{http://arxiv.org/abs/2002.06381}{{\ttfamily arXiv:2002.06381
  [astro-ph.CO]}}.

\bibitem{Barker:2020gcp}
W.~E.~V. Barker, A.~N. Lasenby, M.~P. Hobson, and W.~J. Handley, ``{Systematic
  study of background cosmology in unitary Poincar\'e gauge theories with
  application to emergent dark radiation and $H_0$ tension},''
  \href{http://dx.doi.org/10.1103/PhysRevD.102.024048}{{\em Phys. Rev. D}
  {\bfseries 102} no.~2, (2020) 024048},
  \href{http://arxiv.org/abs/2003.02690}{{\ttfamily arXiv:2003.02690 [gr-qc]}}.

\bibitem{Wang:2020zfv}
D.~Wang and D.~Mota, ``{Can $f(T)$ gravity resolve the $H_0$ tension?},''
  \href{http://dx.doi.org/10.1103/PhysRevD.102.063530}{{\em Phys. Rev. D}
  {\bfseries 102} no.~6, (2020) 063530},
  \href{http://arxiv.org/abs/2003.10095}{{\ttfamily arXiv:2003.10095
  [astro-ph.CO]}}.

\bibitem{Ballardini:2020iws}
M.~Ballardini, M.~Braglia, F.~Finelli, D.~Paoletti, A.~A. Starobinsky, and
  C.~Umilt\`a, ``{Scalar-tensor theories of gravity, neutrino physics, and the
  $H_0$ tension},'' \href{http://dx.doi.org/10.1088/1475-7516/2020/10/044}{{\em
  JCAP} {\bfseries 10} (2020) 044},
  \href{http://arxiv.org/abs/2004.14349}{{\ttfamily arXiv:2004.14349
  [astro-ph.CO]}}.

\bibitem{LinaresCedeno:2020uxx}
F.~X. Linares Cede\~no and U.~Nucamendi, ``{Revisiting cosmological diffusion
  models in Unimodular Gravity and the $H_0$ tension},''
  \href{http://dx.doi.org/10.1016/j.dark.2021.100807}{{\em Phys. Dark Univ.}
  {\bfseries 32} (2021) 100807},
  \href{http://arxiv.org/abs/2009.10268}{{\ttfamily arXiv:2009.10268
  [astro-ph.CO]}}.

\bibitem{daSilva:2020bdc}
W.~J.~C. da~Silva and R.~Silva, ``{Cosmological Perturbations in the Tsallis
  Holographic Dark Energy Scenarios},''
  \href{http://dx.doi.org/10.1140/epjp/s13360-021-01522-9}{{\em Eur. Phys. J.
  Plus} {\bfseries 136} no.~5, (2021) 543},
  \href{http://arxiv.org/abs/2011.09520}{{\ttfamily arXiv:2011.09520
  [astro-ph.CO]}}.

\bibitem{Odintsov:2020qzd}
S.~D. Odintsov, D.~S\'aez-Chill\'on~G\'omez, and G.~S. Sharov, ``{Analyzing the
  $H_0$ tension in $F(R)$ gravity models},''
  \href{http://dx.doi.org/10.1016/j.nuclphysb.2021.115377}{{\em Nucl. Phys. B}
  {\bfseries 966} (2021) 115377},
  \href{http://arxiv.org/abs/2011.03957}{{\ttfamily arXiv:2011.03957 [gr-qc]}}.

\bibitem{Abdalla:2022yfr}
E.~Abdalla {\em et~al.}, ``{Cosmology intertwined: A review of the particle
  physics, astrophysics, and cosmology associated with the cosmological
  tensions and anomalies},''
  \href{http://dx.doi.org/10.1016/j.jheap.2022.04.002}{{\em JHEAp} {\bfseries
  34} (2022) 49--211}, \href{http://arxiv.org/abs/2203.06142}{{\ttfamily
  arXiv:2203.06142 [astro-ph.CO]}}.

\bibitem{CANTATA:2021ktz}
{\bfseries CANTATA} Collaboration, E.~N. Saridakis {\em et~al.}, ``{Modified
  Gravity and Cosmology: An Update by the CANTATA Network},''
  \href{http://arxiv.org/abs/2105.12582}{{\ttfamily arXiv:2105.12582 [gr-qc]}}.

\bibitem{Capozziello:2011et}
S.~Capozziello and M.~De~Laurentis, ``{Extended Theories of Gravity},''
  \href{http://dx.doi.org/10.1016/j.physrep.2011.09.003}{{\em Phys. Rept.}
  {\bfseries 509} (2011) 167--321},
  \href{http://arxiv.org/abs/1108.6266}{{\ttfamily arXiv:1108.6266 [gr-qc]}}.

\bibitem{DeFelice:2010aj}
A.~De~Felice and S.~Tsujikawa, ``{f(R) theories},''
  \href{http://dx.doi.org/10.12942/lrr-2010-3}{{\em Living Rev. Rel.}
  {\bfseries 13} (2010) 3}, \href{http://arxiv.org/abs/1002.4928}{{\ttfamily
  arXiv:1002.4928 [gr-qc]}}.

\bibitem{Nojiri:2010wj}
S.~Nojiri and S.~D. Odintsov, ``{Unified cosmic history in modified gravity:
  from F(R) theory to Lorentz non-invariant models},''
  \href{http://dx.doi.org/10.1016/j.physrep.2011.04.001}{{\em Phys. Rept.}
  {\bfseries 505} (2011) 59--144},
  \href{http://arxiv.org/abs/1011.0544}{{\ttfamily arXiv:1011.0544 [gr-qc]}}.

\bibitem{Starobinsky:2007hu}
A.~A. Starobinsky, ``{Disappearing cosmological constant in f(R) gravity},''
  \href{http://dx.doi.org/10.1134/S0021364007150027}{{\em JETP Lett.}
  {\bfseries 86} (2007) 157--163},
  \href{http://arxiv.org/abs/0706.2041}{{\ttfamily arXiv:0706.2041
  [astro-ph]}}.

\bibitem{Cognola:2007zu}
G.~Cognola, E.~Elizalde, S.~Nojiri, S.~D. Odintsov, L.~Sebastiani, and
  S.~Zerbini, ``{A Class of viable modified f(R) gravities describing inflation
  and the onset of accelerated expansion},''
  \href{http://dx.doi.org/10.1103/PhysRevD.77.046009}{{\em Phys. Rev. D}
  {\bfseries 77} (2008) 046009},
  \href{http://arxiv.org/abs/0712.4017}{{\ttfamily arXiv:0712.4017 [hep-th]}}.

\bibitem{Amendola:2007nt}
L.~Amendola and S.~Tsujikawa, ``{Phantom crossing, equation-of-state
  singularities, and local gravity constraints in f(R) models},''
  \href{http://dx.doi.org/10.1016/j.physletb.2007.12.041}{{\em Phys. Lett. B}
  {\bfseries 660} (2008) 125--132},
  \href{http://arxiv.org/abs/0705.0396}{{\ttfamily arXiv:0705.0396
  [astro-ph]}}.

\bibitem{delaCruz-Dombriz:2006kob}
A.~de~la Cruz-Dombriz and A.~Dobado, ``{A f(R) gravity without cosmological
  constant},'' \href{http://dx.doi.org/10.1103/PhysRevD.74.087501}{{\em Phys.
  Rev. D} {\bfseries 74} (2006) 087501},
  \href{http://arxiv.org/abs/gr-qc/0607118}{{\ttfamily arXiv:gr-qc/0607118}}.

\bibitem{Zhang:2005vt}
P.~Zhang, ``{Testing $f(R)$ gravity against the large scale structure of the
  universe.},'' \href{http://dx.doi.org/10.1103/PhysRevD.73.123504}{{\em Phys.
  Rev. D} {\bfseries 73} (2006) 123504},
  \href{http://arxiv.org/abs/astro-ph/0511218}{{\ttfamily
  arXiv:astro-ph/0511218}}.

\bibitem{Faraoni:2007yn}
V.~Faraoni, ``{de Sitter space and the equivalence between f(R) and
  scalar-tensor gravity},''
  \href{http://dx.doi.org/10.1103/PhysRevD.75.067302}{{\em Phys. Rev. D}
  {\bfseries 75} (2007) 067302},
  \href{http://arxiv.org/abs/gr-qc/0703044}{{\ttfamily arXiv:gr-qc/0703044}}.

\bibitem{Basilakos:2013nfa}
S.~Basilakos, S.~Nesseris, and L.~Perivolaropoulos, ``{Observational
  constraints on viable f(R) parametrizations with geometrical and dynamical
  probes},'' \href{http://dx.doi.org/10.1103/PhysRevD.87.123529}{{\em Phys.
  Rev. D} {\bfseries 87} no.~12, (2013) 123529},
  \href{http://arxiv.org/abs/1302.6051}{{\ttfamily arXiv:1302.6051
  [astro-ph.CO]}}.

\bibitem{Papanikolaou:2021uhe}
T.~Papanikolaou, C.~Tzerefos, S.~Basilakos, and E.~N. Saridakis, ``{Scalar
  induced gravitational waves from primordial black hole Poisson fluctuations
  in f(R) gravity},''
  \href{http://dx.doi.org/10.1088/1475-7516/2022/10/013}{{\em JCAP} {\bfseries
  10} (2022) 013}, \href{http://arxiv.org/abs/2112.15059}{{\ttfamily
  arXiv:2112.15059 [astro-ph.CO]}}.

\bibitem{Nojiri:2005jg}
S.~Nojiri and S.~D. Odintsov, ``{Modified Gauss-Bonnet theory as gravitational
  alternative for dark energy},''
  \href{http://dx.doi.org/10.1016/j.physletb.2005.10.010}{{\em Phys. Lett. B}
  {\bfseries 631} (2005) 1--6},
  \href{http://arxiv.org/abs/hep-th/0508049}{{\ttfamily arXiv:hep-th/0508049}}.

\bibitem{DeFelice:2008wz}
A.~De~Felice and S.~Tsujikawa, ``{Construction of cosmologically viable f(G)
  dark energy models},''
  \href{http://dx.doi.org/10.1016/j.physletb.2009.03.060}{{\em Phys. Lett. B}
  {\bfseries 675} (2009) 1--8},
  \href{http://arxiv.org/abs/0810.5712}{{\ttfamily arXiv:0810.5712 [hep-th]}}.

\bibitem{Zhao:2012vta}
Y.-Y. Zhao, Y.-B. Wu, J.-B. Lu, Z.~Zhang, W.-L. Han, and L.-L. Lin, ``{Modified
  f(G) gravity models with curvature-matter coupling},''
  \href{http://dx.doi.org/10.1140/epjc/s10052-012-1924-2}{{\em Eur. Phys. J. C}
  {\bfseries 72} (2012) 1924}, \href{http://arxiv.org/abs/1203.5593}{{\ttfamily
  arXiv:1203.5593 [astro-ph.CO]}}.

\bibitem{Shamir:2020ckh}
M.~F. Shamir and T.~Naz, ``{Stellar structures in f(G) gravity admitting
  Noether symmetries},''
  \href{http://dx.doi.org/10.1016/j.physletb.2020.135519}{{\em Phys. Lett. B}
  {\bfseries 806} (2020) 135519},
  \href{http://arxiv.org/abs/2006.03339}{{\ttfamily arXiv:2006.03339 [gr-qc]}}.

\bibitem{Asimakis:2022mbe}
P.~Asimakis, S.~Basilakos, and E.~N. Saridakis, ``{Building cubic gravity with
  healthy and viable scalar and tensor perturbations},''
  \href{http://arxiv.org/abs/2212.12494}{{\ttfamily arXiv:2212.12494 [gr-qc]}}.

\bibitem{Lovelock:1971yv}
D.~Lovelock, ``{The Einstein tensor and its generalizations},''
  \href{http://dx.doi.org/10.1063/1.1665613}{{\em J. Math. Phys.} {\bfseries
  12} (1971) 498--501}.

\bibitem{Deruelle:1989fj}
N.~Deruelle and L.~Farina-Busto, ``{The Lovelock Gravitational Field Equations
  in Cosmology},'' \href{http://dx.doi.org/10.1103/PhysRevD.41.3696}{{\em Phys.
  Rev. D} {\bfseries 41} (1990) 3696}.

\bibitem{Cai:2015emx}
Y.-F. Cai, S.~Capozziello, M.~De~Laurentis, and E.~N. Saridakis, ``{f(T)
  teleparallel gravity and cosmology},''
  \href{http://dx.doi.org/10.1088/0034-4885/79/10/106901}{{\em Rept. Prog.
  Phys.} {\bfseries 79} no.~10, (2016) 106901},
  \href{http://arxiv.org/abs/1511.07586}{{\ttfamily arXiv:1511.07586 [gr-qc]}}.

\bibitem{Bengochea:2008gz}
G.~R. Bengochea and R.~Ferraro, ``{Dark torsion as the cosmic speed-up},''
  \href{http://dx.doi.org/10.1103/PhysRevD.79.124019}{{\em Phys. Rev. D}
  {\bfseries 79} (2009) 124019},
  \href{http://arxiv.org/abs/0812.1205}{{\ttfamily arXiv:0812.1205
  [astro-ph]}}.

\bibitem{Linder:2010py}
E.~V. Linder, ``{Einstein's Other Gravity and the Acceleration of the
  Universe},'' \href{http://dx.doi.org/10.1103/PhysRevD.81.127301}{{\em Phys.
  Rev. D} {\bfseries 81} (2010) 127301},
  \href{http://arxiv.org/abs/1005.3039}{{\ttfamily arXiv:1005.3039
  [astro-ph.CO]}}. [Erratum: Phys.Rev.D 82, 109902 (2010)].

\bibitem{Chen:2010va}
S.-H. Chen, J.~B. Dent, S.~Dutta, and E.~N. Saridakis, ``{Cosmological
  perturbations in f(T) gravity},''
  \href{http://dx.doi.org/10.1103/PhysRevD.83.023508}{{\em Phys. Rev. D}
  {\bfseries 83} (2011) 023508},
  \href{http://arxiv.org/abs/1008.1250}{{\ttfamily arXiv:1008.1250
  [astro-ph.CO]}}.

\bibitem{Tamanini:2012hg}
N.~Tamanini and C.~G. Boehmer, ``{Good and bad tetrads in f(T) gravity},''
  \href{http://dx.doi.org/10.1103/PhysRevD.86.044009}{{\em Phys. Rev. D}
  {\bfseries 86} (2012) 044009},
  \href{http://arxiv.org/abs/1204.4593}{{\ttfamily arXiv:1204.4593 [gr-qc]}}.

\bibitem{Bengochea:2010sg}
G.~R. Bengochea, ``{Observational information for f(T) theories and Dark
  Torsion},'' \href{http://dx.doi.org/10.1016/j.physletb.2010.11.064}{{\em
  Phys. Lett. B} {\bfseries 695} (2011) 405--411},
  \href{http://arxiv.org/abs/1008.3188}{{\ttfamily arXiv:1008.3188
  [astro-ph.CO]}}.

\bibitem{Liu:2012fk}
D.~Liu and M.~J. Reboucas, ``{Energy conditions bounds on f(T) gravity},''
  \href{http://dx.doi.org/10.1103/PhysRevD.86.083515}{{\em Phys. Rev. D}
  {\bfseries 86} (2012) 083515},
  \href{http://arxiv.org/abs/1207.1503}{{\ttfamily arXiv:1207.1503
  [astro-ph.CO]}}.

\bibitem{Daouda:2012nj}
M.~H. Daouda, M.~E. Rodrigues, and M.~J.~S. Houndjo, ``{Anisotropic fluid for a
  set of non-diagonal tetrads in f(T) gravity},''
  \href{http://dx.doi.org/10.1016/j.physletb.2012.07.039}{{\em Phys. Lett. B}
  {\bfseries 715} (2012) 241--245},
  \href{http://arxiv.org/abs/1202.1147}{{\ttfamily arXiv:1202.1147 [gr-qc]}}.

\bibitem{MohseniSadjadi:2012brg}
H.~Mohseni~Sadjadi, ``{Generalized Noether symmetry in f(T) gravity},''
  \href{http://dx.doi.org/10.1016/j.physletb.2012.10.073}{{\em Phys. Lett. B}
  {\bfseries 718} (2012) 270--275},
  \href{http://arxiv.org/abs/1210.0937}{{\ttfamily arXiv:1210.0937 [gr-qc]}}.

\bibitem{Finch:2018gkh}
A.~Finch and J.~L. Said, ``{Galactic Rotation Dynamics in f(T) gravity},''
  \href{http://dx.doi.org/10.1140/epjc/s10052-018-6028-1}{{\em Eur. Phys. J. C}
  {\bfseries 78} no.~7, (2018) 560},
  \href{http://arxiv.org/abs/1806.09677}{{\ttfamily arXiv:1806.09677
  [astro-ph.GA]}}.

\bibitem{Golovnev:2020las}
A.~Golovnev and M.-J. Guzm\'an, ``{Bianchi identities in $f (T)$ gravity:
  Paving the way to confrontation with astrophysics},''
  \href{http://dx.doi.org/10.1016/j.physletb.2020.135806}{{\em Phys. Lett. B}
  {\bfseries 810} (2020) 135806},
  \href{http://arxiv.org/abs/2006.08507}{{\ttfamily arXiv:2006.08507 [gr-qc]}}.

\bibitem{Bejarano:2014bca}
C.~Bejarano, R.~Ferraro, and M.~J. Guzm\'an, ``{Kerr geometry in f(T)
  gravity},'' \href{http://dx.doi.org/10.1140/epjc/s10052-015-3288-x}{{\em Eur.
  Phys. J. C} {\bfseries 75} (2015) 77},
  \href{http://arxiv.org/abs/1412.0641}{{\ttfamily arXiv:1412.0641 [gr-qc]}}.

\bibitem{Darabi:2019qpz}
F.~Darabi and K.~Atazadeh, ``{f(T) quantum cosmology},''
  \href{http://dx.doi.org/10.1103/PhysRevD.100.023546}{{\em Phys. Rev. D}
  {\bfseries 100} no.~2, (2019) 023546},
  \href{http://arxiv.org/abs/1903.03409}{{\ttfamily arXiv:1903.03409 [gr-qc]}}.

\bibitem{Sahlu:2019jmy}
S.~Sahlu, J.~Ntahompagaze, M.~Elmardi, and A.~Abebe, ``{The Chaplygin gas as a
  model for modified teleparallel gravity?},''
  \href{http://dx.doi.org/10.1140/epjc/s10052-019-7226-1}{{\em Eur. Phys. J. C}
  {\bfseries 79} no.~9, (2019) 749},
  \href{http://arxiv.org/abs/1904.09897}{{\ttfamily arXiv:1904.09897 [gr-qc]}}.

\bibitem{Benetti:2020hxp}
M.~Benetti, S.~Capozziello, and G.~Lambiase, ``{Updating constraints on f(T)
  teleparallel cosmology and the consistency with Big Bang Nucleosynthesis},''
  \href{http://dx.doi.org/10.1093/mnras/staa3368}{{\em Mon. Not. Roy. Astron.
  Soc.} {\bfseries 500} no.~2, (2020) 1795--1805},
  \href{http://arxiv.org/abs/2006.15335}{{\ttfamily arXiv:2006.15335
  [astro-ph.CO]}}.

\bibitem{Golovnev:2021htv}
A.~Golovnev and M.-J. Guzm\'an, ``{Approaches to spherically symmetric
  solutions in f(T) gravity},''
  \href{http://dx.doi.org/10.3390/universe7050121}{{\em Universe} {\bfseries 7}
  no.~5, (2021) 121}, \href{http://arxiv.org/abs/2103.16970}{{\ttfamily
  arXiv:2103.16970 [gr-qc]}}.

\bibitem{Duchaniya:2022rqu}
L.~K. Duchaniya, S.~V. Lohakare, B.~Mishra, and S.~K. Tripathy, ``{Dynamical
  stability analysis of accelerating f(T) gravity models},''
  \href{http://dx.doi.org/10.1140/epjc/s10052-022-10406-w}{{\em Eur. Phys. J.
  C} {\bfseries 82} no.~5, (2022) 448},
  \href{http://arxiv.org/abs/2202.08150}{{\ttfamily arXiv:2202.08150 [gr-qc]}}.

\bibitem{Kofinas:2014owa}
G.~Kofinas and E.~N. Saridakis, ``{Teleparallel equivalent of Gauss-Bonnet
  gravity and its modifications},''
  \href{http://dx.doi.org/10.1103/PhysRevD.90.084044}{{\em Phys. Rev. D}
  {\bfseries 90} (2014) 084044},
  \href{http://arxiv.org/abs/1404.2249}{{\ttfamily arXiv:1404.2249 [gr-qc]}}.

\bibitem{Kofinas:2014daa}
G.~Kofinas and E.~N. Saridakis, ``{Cosmological applications of $F(T,T_G)$
  gravity},'' \href{http://dx.doi.org/10.1103/PhysRevD.90.084045}{{\em Phys.
  Rev. D} {\bfseries 90} (2014) 084045},
  \href{http://arxiv.org/abs/1408.0107}{{\ttfamily arXiv:1408.0107 [gr-qc]}}.

\bibitem{Bahamonde:2015zma}
S.~Bahamonde, C.~G. B\"ohmer, and M.~Wright, ``{Modified teleparallel theories
  of gravity},'' \href{http://dx.doi.org/10.1103/PhysRevD.92.104042}{{\em Phys.
  Rev. D} {\bfseries 92} no.~10, (2015) 104042},
  \href{http://arxiv.org/abs/1508.05120}{{\ttfamily arXiv:1508.05120 [gr-qc]}}.

\bibitem{Farrugia:2018gyz}
G.~Farrugia, J.~Levi~Said, V.~Gakis, and E.~N. Saridakis, ``{Gravitational
  Waves in Modified Teleparallel Theories},''
  \href{http://dx.doi.org/10.1103/PhysRevD.97.124064}{{\em Phys. Rev. D}
  {\bfseries 97} no.~12, (2018) 124064},
  \href{http://arxiv.org/abs/1804.07365}{{\ttfamily arXiv:1804.07365 [gr-qc]}}.

\bibitem{Escamilla-Rivera:2019ulu}
C.~Escamilla-Rivera and J.~Levi~Said, ``{Cosmological viable models in $f(T,B)$
  theory as solutions to the $H_0$ tension},''
  \href{http://dx.doi.org/10.1088/1361-6382/ab939c}{{\em Class. Quant. Grav.}
  {\bfseries 37} no.~16, (2020) 165002},
  \href{http://arxiv.org/abs/1909.10328}{{\ttfamily arXiv:1909.10328 [gr-qc]}}.

\bibitem{Caruana:2020szx}
M.~Caruana, G.~Farrugia, and J.~Levi~Said, ``{Cosmological bouncing solutions
  in $f(T,B)$ gravity},''
  \href{http://dx.doi.org/10.1140/S10052-020-8204-3}{{\em Eur. Phys. J. C}
  {\bfseries 80} no.~7, (2020) 640},
  \href{http://arxiv.org/abs/2007.09925}{{\ttfamily arXiv:2007.09925 [gr-qc]}}.

\bibitem{Moreira:2021xfe}
A.~R.~P. Moreira, J.~E.~G. Silva, F.~C.~E. Lima, and C.~A.~S. Almeida, ``{Thick
  brane in f(T,B) gravity},''
  \href{http://dx.doi.org/10.1103/PhysRevD.103.064046}{{\em Phys. Rev. D}
  {\bfseries 103} no.~6, (2021) 064046},
  \href{http://arxiv.org/abs/2101.10054}{{\ttfamily arXiv:2101.10054
  [hep-th]}}.

\bibitem{Horndeski:1974wa}
G.~W. Horndeski, ``{Second-order scalar-tensor field equations in a
  four-dimensional space},'' \href{http://dx.doi.org/10.1007/BF01807638}{{\em
  Int. J. Theor. Phys.} {\bfseries 10} (1974) 363--384}.

\bibitem{DeFelice:2010nf}
A.~De~Felice and S.~Tsujikawa, ``{Generalized Galileon cosmology},''
  \href{http://dx.doi.org/10.1103/PhysRevD.84.124029}{{\em Phys. Rev. D}
  {\bfseries 84} (2011) 124029},
  \href{http://arxiv.org/abs/1008.4236}{{\ttfamily arXiv:1008.4236 [hep-th]}}.

\bibitem{Deffayet:2011gz}
C.~Deffayet, X.~Gao, D.~A. Steer, and G.~Zahariade, ``{From k-essence to
  generalised Galileons},''
  \href{http://dx.doi.org/10.1103/PhysRevD.84.064039}{{\em Phys. Rev. D}
  {\bfseries 84} (2011) 064039},
  \href{http://arxiv.org/abs/1103.3260}{{\ttfamily arXiv:1103.3260 [hep-th]}}.

\bibitem{Mota:2010bs}
D.~F. Mota, M.~Sandstad, and T.~Zlosnik, ``{Cosmology of the selfaccelerating
  third order Galileon},''
  \href{http://dx.doi.org/10.1007/JHEP12(2010)051}{{\em JHEP} {\bfseries 12}
  (2010) 051}, \href{http://arxiv.org/abs/1009.6151}{{\ttfamily arXiv:1009.6151
  [astro-ph.CO]}}.

\bibitem{Barreira:2013eea}
A.~Barreira, B.~Li, W.~A. Hellwing, C.~M. Baugh, and S.~Pascoli, ``{Nonlinear
  structure formation in the Cubic Galileon gravity model},''
  \href{http://dx.doi.org/10.1088/1475-7516/2013/10/027}{{\em JCAP} {\bfseries
  10} (2013) 027}, \href{http://arxiv.org/abs/1306.3219}{{\ttfamily
  arXiv:1306.3219 [astro-ph.CO]}}.

\bibitem{Qiu:2011cy}
T.~Qiu, J.~Evslin, Y.-F. Cai, M.~Li, and X.~Zhang, ``{Bouncing Galileon
  Cosmologies},'' \href{http://dx.doi.org/10.1088/1475-7516/2011/10/036}{{\em
  JCAP} {\bfseries 10} (2011) 036},
  \href{http://arxiv.org/abs/1108.0593}{{\ttfamily arXiv:1108.0593 [hep-th]}}.

\bibitem{Appleby:2012ba}
S.~A. Appleby and E.~V. Linder, ``{Trial of Galileon gravity by cosmological
  expansion and growth observations},''
  \href{http://dx.doi.org/10.1088/1475-7516/2012/08/026}{{\em JCAP} {\bfseries
  08} (2012) 026}, \href{http://arxiv.org/abs/1204.4314}{{\ttfamily
  arXiv:1204.4314 [astro-ph.CO]}}.

\bibitem{Barreira:2014jha}
A.~Barreira, B.~Li, C.~Baugh, and S.~Pascoli, ``{The observational status of
  Galileon gravity after Planck},''
  \href{http://dx.doi.org/10.1088/1475-7516/2014/08/059}{{\em JCAP} {\bfseries
  08} (2014) 059}, \href{http://arxiv.org/abs/1406.0485}{{\ttfamily
  arXiv:1406.0485 [astro-ph.CO]}}.

\bibitem{Arroja:2015wpa}
F.~Arroja, N.~Bartolo, P.~Karmakar, and S.~Matarrese, ``{The two faces of
  mimetic Horndeski gravity: disformal transformations and Lagrange
  multiplier},'' \href{http://dx.doi.org/10.1088/1475-7516/2015/09/051}{{\em
  JCAP} {\bfseries 09} (2015) 051},
  \href{http://arxiv.org/abs/1506.08575}{{\ttfamily arXiv:1506.08575 [gr-qc]}}.

\bibitem{Hinterbichler:2015pqa}
K.~Hinterbichler and A.~Joyce, ``{Hidden symmetry of the Galileon},''
  \href{http://dx.doi.org/10.1103/PhysRevD.92.023503}{{\em Phys. Rev. D}
  {\bfseries 92} no.~2, (2015) 023503},
  \href{http://arxiv.org/abs/1501.07600}{{\ttfamily arXiv:1501.07600
  [hep-th]}}.

\bibitem{Babichev:2015rva}
E.~Babichev, C.~Charmousis, and M.~Hassaine, ``{Charged Galileon black
  holes},'' \href{http://dx.doi.org/10.1088/1475-7516/2015/05/031}{{\em JCAP}
  {\bfseries 05} (2015) 031}, \href{http://arxiv.org/abs/1503.02545}{{\ttfamily
  arXiv:1503.02545 [gr-qc]}}.

\bibitem{Brax:2011sv}
P.~Brax, C.~Burrage, and A.-C. Davis, ``{Laboratory Tests of the Galileon},''
  \href{http://dx.doi.org/10.1088/1475-7516/2011/09/020}{{\em JCAP} {\bfseries
  09} (2011) 020}, \href{http://arxiv.org/abs/1106.1573}{{\ttfamily
  arXiv:1106.1573 [hep-ph]}}.

\bibitem{Renk:2017rzu}
J.~Renk, M.~Zumalac\'arregui, F.~Montanari, and A.~Barreira, ``{Galileon
  gravity in light of ISW, CMB, BAO and H$_0$ data},''
  \href{http://dx.doi.org/10.1088/1475-7516/2017/10/020}{{\em JCAP} {\bfseries
  10} (2017) 020}, \href{http://arxiv.org/abs/1707.02263}{{\ttfamily
  arXiv:1707.02263 [astro-ph.CO]}}.

\bibitem{Gleyzes:2014dya}
J.~Gleyzes, D.~Langlois, F.~Piazza, and F.~Vernizzi, ``{Healthy theories beyond
  Horndeski},'' \href{http://dx.doi.org/10.1103/PhysRevLett.114.211101}{{\em
  Phys. Rev. Lett.} {\bfseries 114} no.~21, (2015) 211101},
  \href{http://arxiv.org/abs/1404.6495}{{\ttfamily arXiv:1404.6495 [hep-th]}}.

\bibitem{Langlois:2015cwa}
D.~Langlois and K.~Noui, ``{Degenerate higher derivative theories beyond
  Horndeski: evading the Ostrogradski instability},''
  \href{http://dx.doi.org/10.1088/1475-7516/2016/02/034}{{\em JCAP} {\bfseries
  02} (2016) 034}, \href{http://arxiv.org/abs/1510.06930}{{\ttfamily
  arXiv:1510.06930 [gr-qc]}}.

\bibitem{Langlois:2018dxi}
D.~Langlois, ``{Dark energy and modified gravity in degenerate higher-order
  scalar\textendash{}tensor (DHOST) theories: A review},''
  \href{http://dx.doi.org/10.1142/S0218271819420069}{{\em Int. J. Mod. Phys. D}
  {\bfseries 28} no.~05, (2019) 1942006},
  \href{http://arxiv.org/abs/1811.06271}{{\ttfamily arXiv:1811.06271 [gr-qc]}}.

\bibitem{Babichev:2017guv}
E.~Babichev, C.~Charmousis, and A.~Leh\'ebel, ``{Asymptotically flat black
  holes in Horndeski theory and beyond},''
  \href{http://dx.doi.org/10.1088/1475-7516/2017/04/027}{{\em JCAP} {\bfseries
  04} (2017) 027}, \href{http://arxiv.org/abs/1702.01938}{{\ttfamily
  arXiv:1702.01938 [gr-qc]}}.

\bibitem{Ilyas:2020qja}
A.~Ilyas, M.~Zhu, Y.~Zheng, Y.-F. Cai, and E.~N. Saridakis, ``{DHOST Bounce},''
  \href{http://dx.doi.org/10.1088/1475-7516/2020/09/002}{{\em JCAP} {\bfseries
  09} (2020) 002}, \href{http://arxiv.org/abs/2002.08269}{{\ttfamily
  arXiv:2002.08269 [gr-qc]}}.

\bibitem{Naruko:2015zze}
A.~Naruko, D.~Yoshida, and S.~Mukohyama, ``{Gravitational
  scalar\textendash{}tensor theory},''
  \href{http://dx.doi.org/10.1088/0264-9381/33/9/09LT01}{{\em Class. Quant.
  Grav.} {\bfseries 33} no.~9, (2016) 09LT01},
  \href{http://arxiv.org/abs/1512.06977}{{\ttfamily arXiv:1512.06977 [gr-qc]}}.

\bibitem{Saridakis:2016ahq}
E.~N. Saridakis and M.~Tsoukalas, ``{Cosmology in new gravitational
  scalar-tensor theories},''
  \href{http://dx.doi.org/10.1103/PhysRevD.93.124032}{{\em Phys. Rev. D}
  {\bfseries 93} no.~12, (2016) 124032},
  \href{http://arxiv.org/abs/1601.06734}{{\ttfamily arXiv:1601.06734 [gr-qc]}}.

\bibitem{Heisenberg:2022lob}
L.~Heisenberg, H.~Villarrubia-Rojo, and J.~Zosso, ``{Simultaneously solving the
  H0 and \ensuremath{\sigma}8 tensions with late dark energy},''
  \href{http://dx.doi.org/10.1016/j.dark.2022.101163}{{\em Phys. Dark Univ.}
  {\bfseries 39} (2023) 101163},
  \href{http://arxiv.org/abs/2201.11623}{{\ttfamily arXiv:2201.11623
  [astro-ph.CO]}}.

\bibitem{Heisenberg:2022gqk}
L.~Heisenberg, H.~Villarrubia-Rojo, and J.~Zosso, ``{Can late-time extensions
  solve the H0 and \ensuremath{\sigma}8 tensions?},''
  \href{http://dx.doi.org/10.1103/PhysRevD.106.043503}{{\em Phys. Rev. D}
  {\bfseries 106} no.~4, (2022) 043503},
  \href{http://arxiv.org/abs/2202.01202}{{\ttfamily arXiv:2202.01202
  [astro-ph.CO]}}.

\bibitem{DeFelice:2011bh}
A.~De~Felice and S.~Tsujikawa, ``{Conditions for the cosmological viability of
  the most general scalar-tensor theories and their applications to extended
  Galileon dark energy models},''
  \href{http://dx.doi.org/10.1088/1475-7516/2012/02/007}{{\em JCAP} {\bfseries
  02} (2012) 007}, \href{http://arxiv.org/abs/1110.3878}{{\ttfamily
  arXiv:1110.3878 [gr-qc]}}.

\bibitem{Kobayashi:2011nu}
T.~Kobayashi, M.~Yamaguchi, and J.~Yokoyama, ``{Generalized G-inflation:
  Inflation with the most general second-order field equations},''
  \href{http://dx.doi.org/10.1143/PTP.126.511}{{\em Prog. Theor. Phys.}
  {\bfseries 126} (2011) 511--529},
  \href{http://arxiv.org/abs/1105.5723}{{\ttfamily arXiv:1105.5723 [hep-th]}}.

\bibitem{Petronikolou:2021shp}
M.~Petronikolou, S.~Basilakos, and E.~N. Saridakis, ``{Alleviating $H_0$
  tension in Horndeski gravity},''
  \href{http://arxiv.org/abs/2110.01338}{{\ttfamily arXiv:2110.01338 [gr-qc]}}.

\bibitem{Saridakis:2010mf}
E.~N. Saridakis and S.~V. Sushkov, ``{Quintessence and phantom cosmology with
  non-minimal derivative coupling},''
  \href{http://dx.doi.org/10.1103/PhysRevD.81.083510}{{\em Phys. Rev. D}
  {\bfseries 81} (2010) 083510},
  \href{http://arxiv.org/abs/1002.3478}{{\ttfamily arXiv:1002.3478 [gr-qc]}}.

\bibitem{Koutsoumbas:2017fxp}
G.~Koutsoumbas, K.~Ntrekis, E.~Papantonopoulos, and E.~N. Saridakis,
  ``{Unification of Dark Matter - Dark Energy in Generalized Galileon
  Theories},'' \href{http://dx.doi.org/10.1088/1475-7516/2018/02/003}{{\em
  JCAP} {\bfseries 02} (2018) 003},
  \href{http://arxiv.org/abs/1704.08640}{{\ttfamily arXiv:1704.08640 [gr-qc]}}.

\bibitem{Karydas:2021wmx}
S.~Karydas, E.~Papantonopoulos, and E.~N. Saridakis, ``{Successful Higgs
  inflation from combined nonminimal and derivative couplings},''
  \href{http://dx.doi.org/10.1103/PhysRevD.104.023530}{{\em Phys. Rev. D}
  {\bfseries 104} no.~2, (2021) 023530},
  \href{http://arxiv.org/abs/2102.08450}{{\ttfamily arXiv:2102.08450 [gr-qc]}}.

\bibitem{Bellini:2014fua}
E.~Bellini and I.~Sawicki, ``{Maximal freedom at minimum cost: linear
  large-scale structure in general modifications of gravity},''
  \href{http://dx.doi.org/10.1088/1475-7516/2014/07/050}{{\em JCAP} {\bfseries
  07} (2014) 050}, \href{http://arxiv.org/abs/1404.3713}{{\ttfamily
  arXiv:1404.3713 [astro-ph.CO]}}.

\bibitem{Peirone:2017ywi}
S.~Peirone, K.~Koyama, L.~Pogosian, M.~Raveri, and A.~Silvestri, ``{Large-scale
  structure phenomenology of viable Horndeski theories},''
  \href{http://dx.doi.org/10.1103/PhysRevD.97.043519}{{\em Phys. Rev. D}
  {\bfseries 97} no.~4, (2018) 043519},
  \href{http://arxiv.org/abs/1712.00444}{{\ttfamily arXiv:1712.00444
  [astro-ph.CO]}}.

\bibitem{DeFelice:2010pv}
A.~De~Felice and S.~Tsujikawa, ``{Cosmology of a covariant Galileon field},''
  \href{http://dx.doi.org/10.1103/PhysRevLett.105.111301}{{\em Phys. Rev.
  Lett.} {\bfseries 105} (2010) 111301},
  \href{http://arxiv.org/abs/1007.2700}{{\ttfamily arXiv:1007.2700
  [astro-ph.CO]}}.

\bibitem{Appleby:2011aa}
S.~Appleby and E.~V. Linder, ``{The Paths of Gravity in Galileon Cosmology},''
  \href{http://dx.doi.org/10.1088/1475-7516/2012/03/043}{{\em JCAP} {\bfseries
  03} (2012) 043}, \href{http://arxiv.org/abs/1112.1981}{{\ttfamily
  arXiv:1112.1981 [astro-ph.CO]}}.

\bibitem{Ezquiaga:2017ekz}
J.~M. Ezquiaga and M.~Zumalac\'arregui, ``{Dark Energy After GW170817: Dead
  Ends and the Road Ahead},''
  \href{http://dx.doi.org/10.1103/PhysRevLett.119.251304}{{\em Phys. Rev.
  Lett.} {\bfseries 119} no.~25, (2017) 251304},
  \href{http://arxiv.org/abs/1710.05901}{{\ttfamily arXiv:1710.05901
  [astro-ph.CO]}}.

\bibitem{Nesseris:2013jea}
S.~Nesseris, S.~Basilakos, E.~N. Saridakis, and L.~Perivolaropoulos, ``{Viable
  $f(T)$ models are practically indistinguishable from $\Lambda$CDM},''
  \href{http://dx.doi.org/10.1103/PhysRevD.88.103010}{{\em Phys. Rev. D}
  {\bfseries 88} (2013) 103010},
  \href{http://arxiv.org/abs/1308.6142}{{\ttfamily arXiv:1308.6142
  [astro-ph.CO]}}.

\bibitem{Banerjee:2022ynv}
S.~Banerjee, M.~Petronikolou, and E.~N. Saridakis, ``{Alleviating the H0
  tension with new gravitational scalar tensor theories},''
  \href{http://dx.doi.org/10.1103/PhysRevD.108.024012}{{\em Phys. Rev. D}
  {\bfseries 108} no.~2, (2023) 024012},
  \href{http://arxiv.org/abs/2209.02426}{{\ttfamily arXiv:2209.02426 [gr-qc]}}.

\end{thebibliography}\endgroup


\end{document}